\documentclass{svjour2}                    

\smartqed  

\usepackage{graphicx}
\usepackage{epsfig}

\journalname{Journal of Statistical Physics}
\begin{document}

\title{Schwinger Boson Formulation and Solution of the Crow-Kimura
and Eigen Models of Quasispecies Theory}

\author{Jeong-Man Park $^{1,2}$ \and
Michael W.\ Deem$^1$
}

\institute{
Michael W.\ Deem \at
Department of Physics \& Astronomy,
Rice University, Houston,Texas 77005--1892\\
Tel.: 713-348-5852\\
\email{mwdeem@rice.edu}\\
\and
J.-M.\ Park \at
Department of Physics \& Astronomy,
Rice University, Houston,Texas 77005--1892\\
and
Department of Physics,
The Catholic University of Korea, 
Puchon 420--743, Korea\\
\email{jeong@king.rice.edu}\\
}

\date{Received: date / Accepted: date}

\maketitle

\begin{abstract}
We express the 
Crow-Kimura and Eigen models of quasispecies theory
in a functional integral representation.
We formulate the spin coherent state functional integrals
using the Schwinger Boson method.
In this formulation, we are able to deduce the long-time
behavior of these models for arbitrary 
replication and degradation functions.
We discuss the phase transitions that occur in these models
as a function of mutation rate.
We derive for these models the leading
order corrections to the infinite genome length limit.
\end{abstract}

\section{Introduction}
The quasispecies models of Eigen \cite{ei71} and
Crow-Kimura \cite{ck70} are among the simplest that
capture basic aspects of mutation and evolutionary
selection in large, homogeneous populations of viruses.
These models are a favorite entry point for physicists
to evolutionary biology, due to the phase transitions
that they exhibit and their mathematical simplicity
\cite{Krug}.

While the models were originally defined in the continuous
time limit, the first connections to statistical mechanics
were made to the discrete-time versions of these models.
In particular, the discrete-time Eigen model, which entails the
additional assumption of allowing only a single mutation at
one point in time in addition to discretization of time, was 
shown to be equivalent to a particular type of
Ising model \cite{Leuthausser}.  
A distinction between
the bulk magnetization and the observable, surface, magnetization
was discovered in the discrete-time Eigen
model \cite{Tarazona}, and an analogy to
the surface wetting phenomenon in condensed matter was made.
\emph{Magnetization} is a now-standard term in the physical
quasispecies field for average composition.  That is, the magnetization
at site $j$ on the genome is the average composition of the
base at that site, averaged over all sequences in the population.
  A functional
integral representation of the discrete-time Eigen model
was introduced through use of functional delta functions
\cite{Peliti2002}.  With this representation, solution of
this particular model was possible for viral
replication rates that depend in an arbitrary way
on distance from a single point in genome space.
The closely related Crow-Kimura, or parallel or para-muse, 
continuous-time model was
formulated as a quantum spin Hamiltonian in \cite{bb97,bb98}.
It was formulated as a functional integral and solved for  viral
replication rates that depend in an arbitrary way
on distance from a single point in genome space
in \cite{sh04a,sh04b,sh04c}.
The discrete-time Eigen model in a sense interpolates
between the continuous-time Eigen model and the Crow-Kimura 
model, because its limit as the time step becomes
small is the Crow-Kimura model rather than the
Eigen model.  What distinguishes the continuous-time
Eigen model from the other models is the possibility of multiple
mutation events in an infinitesimal time step \cite{Krug}.

In this manuscript we seek to provide a detailed derivation for
a functional integral representation of the continuous-time 
parallel and Eigen models of quasispecies theory. 
 These representations
are used to find exact solutions of these continuous-time
 quasispecies theories in the
limit of large genomes.  
These results are used to exhibit the phase transitions
that occur in these models as a function of mutation rate.
The coherent states formalism that we introduce
allows for the first time the expression of the full time-dependent
probability distribution in sequence space
of these quasispecies theories as a function of arbitrary initial
and final conditions.
In addition, we use the functional integral
expressions to obtain for the first time the $O(1/N)$ corrections to the mean
replication rates in these models.  Finally, we also 
use the functional integral
expressions to find for the first time the $O(1/\sqrt N)$ width of the
virus populations around the most probable genomes in sequence 
space.

The rest of the manuscript is organized as follows.  In Section
2 we map the continuous-time parallel model onto a spin coherent state path
integral using the Schwinger Boson method.  We evaluate
the theory for a general replication rate.
We find the
corrections to the infinite genome limit.  In Section 3
we map the continuous-time Eigen model onto a
functional integral, again using  the Schwinger Boson method. 
While the functional integral appears more singular than in the
parallel case, due to the presence of multiple mutations
at a single time step in this model, we also solve this
model  for a general replication rate.
We also find the
corrections to the infinite genome limit.  We
discuss the  width of the virus population in genome space in
the large $N$  limit
and correlations in the field theory
in Section 4.  We also find the expression for the full
time-dependent probability distribution as it depends on initial
and final conditions.
We conclude in Section 5.  Much of the detailed
derivations are included in Appendices.

\section{Spin coherent state representation of the parallel model}
\subsection{The parallel model}

In the parallel model, the probability distribution of
viruses in the space of all possible viral genomes is considered.
For simplicity, it is assumed that the genome can be written
as a sequence of $N$ binary digits, or spins:
$s_n^i=\pm 1$, $1 \le n \le N$.
Distances in the genome space are calculated by
the Hamming measure:
$d_{ij}=(N-\sum_n s_n^i s_n^j)/2$.
The probability for a virus to be in a given genome
state, $p_i$, $1 \le i \le 2^N$,  satisfies
the parallel model differential equation
\begin{eqnarray}
 \frac{{dp}_i}{dt}={p_i}(r_i-{\sum}_{j=1}^{2^N}r_j
p_j)+{\sum}_{j=1}^{2^N}\mu_{i j}p_j \ .
\label{1}
\end{eqnarray}
Here $r_i$ is the number of offspring per
unit period of time, or
replication rate, and
$\mu_{ij}=\mu \Delta(d_{ij}-1) - N \mu \Delta(d_{ij})$
is the mutation rate to move
from sequence $s_i$ to sequence
$s_j$ per unit period of time.
Here $\Delta(n)$ is the Kronecker delta.
The non-linear term in Eq.\ (\ref{1}) serves
simply to enforce the conservation of
probability, $\sum_i p_i = 1$.  We can express the
differential equation in a simpler, linear form
\begin{eqnarray}
 \frac{{dq}_i}{dt}= r_i q_i
+ {\sum}_{j=1}^{2^N} \mu_{i j} q_j
\label{2}
\end{eqnarray}
with the transformation $p_i(t) = q_i(t) / \sum_j q_j(t)$.
The explicit form of the replication rate is
$r_i = N f(u)$, where $u = (1/N) \sum_n s^i_n$.

\subsection{The parallel model in operator form}
Motivated by the observation \cite{bb97} that the parallel
dynamics in Eq.\ (\ref{2}) is equivalent to quantum
dynamics in imaginary time, we express the model in
an operator form.  We define two kinds of creation and
annihilation operators: $\hat a_\alpha(j), \hat a^\dagger_\alpha(j)$,
$\alpha = 1,2$ and $j = 1, \ldots, N$.  
These operators obey the commutation relations
\begin{eqnarray}
\left[\hat a_\alpha(i) ,  \hat a^\dagger_\beta(j)\right ]
    &=& \delta_{\alpha \beta}
\delta_{i j}
 \nonumber \\ 
\left[\hat a_\alpha(i) ,  \hat a_\beta(j)\right] &=& 0
 \nonumber \\ 
\left[\hat a^\dagger_\alpha(i) ,  \hat a^\dagger_\beta(j)\right] &=& 0
\label{3}
\end{eqnarray}
These operators
create either a spin-up state for $\alpha = 1$
or a spin-down state for $\alpha = 2$ at position $j$ in the
genome.  While it might see more natural to introduce a single
set of creation and annihilation operators to define whether
the spin at position $j$ is up or down, this approach leads to
a non-Gaussian field theory even for a vanishing replication rate
function.  Use of two sets of creation and annihilation
operators leads to a Gaussian field theory, with non-quadratic terms
stemming from the replication rate function.  We find this second form
of the theory more convenient for calculation.  This second form, moreover,
can be extended to the case where the sequence
alphabet is larger than binary.
Since the state is  one and only one of the possible
letters at position $j$, spin-up or spin-down in the binary alphabet case
considered here,
 we will enforce the constraint that
\begin{eqnarray}
\sum_\alpha \hat a^\dagger_\alpha(j) \hat a_\alpha(j) = 1
\label{3a}
\end{eqnarray}
for all $j$.  Thus, the state at site $j$ is either
\begin{eqnarray}
 \vert 1,0 \rangle = [\hat a^\dagger_1(j)]^1 [\hat a^\dagger_2(j)]^0
\vert 0,0 \rangle {\rm ~or~}
\vert 0,1 \rangle = [\hat a^\dagger_1(j)]^0 [\hat a^\dagger_2(j)]^1
\vert 0,0 \rangle
\label{4}
\end{eqnarray}
Defining $n_j^i$ to be the power on $\hat a^\dagger_1(j)$
for spin state $i$, we can rewrite the parallel model
dynamics as
\begin{eqnarray}
\frac{d}{dt} P( \{ n^i_j \})
&=&
r_i( \{ n^i_j \}) P( \{ n^i_j \})
+ \mu \sum_{j=1}^N \bigg[
(1-n_j^i)P( \{ \ldots, n_j^i+1, \ldots \})
\nonumber \\ &&
+ n_j^i P( \{ \ldots, n^i_j-1, \ldots \})
- P( \{ n^i_j \})
\bigg]
\end{eqnarray}
We introduce the state vector 
\begin{eqnarray}
\vert \psi \rangle = \sum_{i=1}^{2^N}
P( \{ n^i \}) \vert \{ n^i \} \rangle
\label{5}
\end{eqnarray}
which satisfies the differential equation
\begin{eqnarray}
\frac{d}{dt} \vert \psi \rangle = \sum_{i=1}^{2^N} 
\frac {d P( \{ n^i \}) } {dt}  \vert \{ n^i \} \rangle
\label{6}
\end{eqnarray}
We now write this Eq.\ (\ref{6}) in operator form.
First, we introduce vector notation for the creation
and annihilation operators: 
$\hat {\vec a}(j) = ( \hat a_1(j), \hat a_2(j))$
and 
$\hat {\vec a}^\dagger (j) =
    ( \hat a_1^\dagger (j), \hat a_2^\dagger (j))$.
Then we introduce operators
\begin{eqnarray} 
T_i(j) &=& \hat {\vec a}^\dagger (j) \sigma_i \hat {\vec a}(j)
\nonumber \\
T_0 &=& \hat {\vec a}^\dagger (j) \cdot \hat {\vec a}(j)
\label{7}
\end{eqnarray}
with spin matrices
\begin{eqnarray} 
\sigma_1 = \left( \begin{array}{cc} 0&1 \\ 1&0 \end{array} \right) ,~~
\sigma_2 = \left( \begin{array}{cc} 0&-1 \\ 1&0 \end{array} \right) ,~~
\sigma_3 = \left( \begin{array}{cc} 1&0 \\ 0&-1 \end{array} \right)
\label{8}
\end{eqnarray}
Then the dynamics of Eq.\ (\ref{6})
can be written as
\begin{eqnarray}
\frac{d}{dt} \vert \psi \rangle
= -\hat H \vert \psi \rangle
\label{9}
\end{eqnarray}
with
\begin{eqnarray}
-\hat H = N f\left[\sum_{j=1}^N T_3(j)/N \right] +
 \mu \sum_{j=1}^N [T_1(j) - T_0(j)]
\label{10}
\end{eqnarray}

\subsection{The field theoretic representation of the parallel model}

We convert this operator form of the parallel model
into a functional integral by using coherent states.
We define a spin coherent state by
\begin{eqnarray}
\vert \vec z(j) \rangle &=&
e^{\hat {\vec a}^\dagger (j) \cdot \vec z(j) -
\vec z^*(j) \cdot \hat {\vec a}(j) }
\vert (0,0)(j) \rangle
\nonumber \\ &=&
e^{-\frac{1}{2} \vec z^*(j) \cdot \vec z(j)}
\sum_{n,m = 0}^\infty \frac{ 
[z_1(j)]^n [z_2(j)]^m }{\sqrt{n! m!}}
\vert (n,m)(j) \rangle
\label{11}
\end{eqnarray}
These coherent states satisfy a completeness relation
\begin{eqnarray}
I = \int \prod_{j=1}^N  \frac{d \vec z^*(j) d \vec z(j) }
{\pi^2} \vert \{ \vert \vec z \} \rangle
\langle \{ \vert \vec z \} \vert
\end{eqnarray}
The overlap of coherent states satisfies
\begin{equation}
\langle \vec z'(j) \vert \vec z(j) \rangle = 
e^{- \frac{1}{2}
\{ {\vec z'}^* (j) \cdot [\vec z'(j)- \vec z(j)] -
   [\vec z'^* (j) - \vec z^*(j)] \cdot \vec z(j) \} }
\end{equation}
Equation (\ref{9})  enforces constraint (\ref{3a}) if the
initial conditions obey the constraint.  To project
arbitrary initial conditions onto this constraint, we
use the operator
\begin{eqnarray}
\hat P &=& \prod_{j=1}^N \hat P(j) = 
\prod_{j=1}^N \Delta[\hat {\vec a}^\dagger (j) \cdot \hat {\vec a}(j) -1]
\nonumber \\ &=&
\int_0^{2 \pi} \prod_{j=1}^N \frac{d \lambda_j}{2 \pi} 
e^{i \lambda_j [\hat {\vec a}^\dagger (j) \cdot \hat {\vec a}(j) -1]}
\label{12}
\end{eqnarray}
The probability to be in a given final state at time $t$ is
\begin{eqnarray}
P( \{ \vec n \}, t) = 
\sum_{\{ \vec n_0 \} }
\langle \{ \vec n \}
\vert e^{-\hat H t} \vert \{ \vec n_0 \}\rangle
P(\{ \vec n_0 \})
\label{13}
\end{eqnarray}
Using the coherent states identity in a Trotter factorization, we find
\begin{eqnarray}
P( \{ \vec n \}, t) &=&
\sum_{\{ \vec n_0 \} }
\lim_{M \to \infty} \langle \{ \vec n \}  \vert
\int \left[ \prod_{j=1}^N \frac{d \vec z^*_M(j) d \vec z_M(j) }
{\pi^2} \right] 
\vert \{  \vec z_M \} \rangle
\langle \{ \vec z_M \} \vert
e^{-\epsilon \hat H}
\nonumber \\ && \times
\int \left[ \prod_{j=1}^N \frac{d \vec z^*_{M-1}(j) d \vec z_{M-1}(j) }
{\pi^2} \right] 
\vert \{ \vec z_{M-1} \} \rangle
\langle \{ \vec z_{M-1} \} \vert
e^{-\epsilon \hat H}
\nonumber \\ && \vdots
\nonumber \\ &&
\int \left[ \prod_{j=1}^N \frac{d \vec z^*_1(j) d \vec z_1(j) }
{\pi^2} \right] 
\vert \{ \vec z_{1} \} \rangle
\langle \{ \vec z_{1} \} \vert
e^{-\epsilon \hat H}
\nonumber \\ && \times
\int \left[ \prod_{j=1}^N \frac{d \vec z^*_0(j) d \vec z_0(j) }
{\pi^2} \right] 
e^{-\epsilon \hat H}
\hat P 
\vert  \{ \vec z_0 \}\rangle
\langle \{ \vec z_0 \} 
\vert  \{ \vec n_0 \}\rangle
P(\{ \vec n_0 \})
\nonumber \\
&=&
\lim_{M \to \infty} \int
[ {\cal D} \vec z^*
{\cal D} \vec z ]
\langle \{ \vec n \} \vert \{ \vec z_M \} \rangle
\left(
\sum_{\{ \vec n_0 \} }
\langle \{ \vec z_0 \} \vert \{ \vec n_0 \} \rangle
P(\{ \vec n_0 \})
\right)
\nonumber \\ &&
\times \prod_{k=2}^M 
\langle \{ \vert \vec z_{k} \} \vert
e^{-\epsilon \hat H}
\vert \{ \vec z_{k-1} \} \rangle
\langle \{ \vec z_{1} \} \vert
e^{-\epsilon \hat H}
\hat P
\vert \{ \vec z_{0} \} \rangle
\label{14}
\end{eqnarray}
For initial conditions that satisfy constraint (\ref{3a})
\begin{eqnarray}
\langle \{ \vec n \} \vert \{ \vec z_M \} \rangle
&=& \prod_j e^{-\frac{1}{2} \vec z^*_M(j) \cdot \vec z_M (j) }
   \vec n(j) \cdot \vec z_M(j) 
\nonumber \\
\langle \{ \vec z_0 \} \vert \{ \vec n_0 \} \rangle
&=& \prod_j e^{-\frac{1}{2} \vec z^*_0(j) \cdot \vec z_0 (j) }
\vec z_0^*(j) \cdot \vec n_0(j) 
\label{16}
\end{eqnarray}
Conversely, if the projection operator is needed for
arbitrary initial conditions, we note
\begin{eqnarray}
\hat P \vert \vec z_0(j) \rangle
&=& \int_0^{2 \pi} \prod_{j=1}^N \frac{d \lambda_j}{2 \pi}
e^{-i \lambda_j} e^{ i \lambda_j \hat {\vec a}^\dagger (j) \cdot
\hat {\vec a}(j) }
\vert \vec z(j) \rangle
\nonumber \\ &=&
 \int_0^{2 \pi} \prod_{j=1}^N \frac{d \lambda_j}{2 \pi}
e^{-i \lambda_j} 
\vert (e^{i \lambda_j} \vec z) (j) \rangle
\label{17}
\end{eqnarray}
For initial conditions that
satisfy constraint (\ref{3a}), we may remove the 
projection operator to find
\begin{eqnarray}
P( \{ \vec n \}, t) &=&
\lim_{M \to \infty} \int
[ {\cal D} \vec z^*
{\cal D} \vec z ]
\sum_{\{ \vec n_0 \} }
P(\{ \vec n_0 \})
\prod_{j =1}^N 
\vec n(j) \cdot \vec z_M(j) 
\vec z_0^*(j) \cdot \vec n_0(j) 
\nonumber \\ && \times
e^{-\frac{1}{2} \vec z^*_M(j) \cdot \vec z_M (j) }
e^{-\frac{1}{2} \vec z^*_0(j) \cdot \vec z_0 (j) }
\nonumber \\ &&
\times \prod_{k =1}^M
e^{-\frac{1}{2}
\{ {\vec z}_k^* (j) \cdot [\vec z_k(j)- \vec z_{k-1}(j)] -
   [(\vec z)^*_k (j) - \vec z^*_{k-1}(j)] \cdot \vec z_{k-1}(j) \} }
\nonumber \\ && \times
e^{
\epsilon N f\left[\sum_{j=1}^N  {\vec z}^*_{k} (j) 
\sigma_3  {\vec z}_{k-1}(j)
/N \right] +
\epsilon    \Delta f({\vec z}^*_{k},  {\vec z}_{k-1})
}
\nonumber \\ && \times e^{ \epsilon  \mu \sum_{j=1}^N [ {\vec z}^*_{k} (j)   
\sigma_1  {\vec z}_{k-1}(j) -1]  }
\nonumber \\
&=&
\lim_{M \to \infty} \int
[ {\cal D} \vec z^*
{\cal D} \vec z ]
\sum_{\{ \vec n_0 \} }
P(\{ \vec n_0 \})
\nonumber \\ && \times
\prod_{j =1}^N 
\vec n(j) \cdot \vec z_M(j) ~
\vec z_0^*(j) \cdot \vec n_0(j) ~ 
e^{-S[\vec z^*, \vec z] }
\label{18}
\end{eqnarray}
where
\begin{eqnarray}
S[\vec z^*, \vec z] &=&
\vec z^*_0(j) \cdot \vec z_0 (j)
+ \sum_{k =1}^M
{\vec z}_k^* (j) \cdot [\vec z_k(j)- \vec z_{k-1}(j)]
\nonumber \\ &&
- \epsilon \sum_{k =1}^M \bigg\{
N f\left[
\sum_{j=1}^N  {\vec z}^*_{k} (j) 
\sigma_3  {\vec z}_{k-1}(j) /N 
\right]
 +  \Delta f({\vec z}^*_{k},  {\vec z}_{k-1})
\nonumber \\ &&
+   \mu \sum_{j=1}^N [ {\vec z}^*_{k} (j)   
\sigma_1  {\vec z}_{k-1}(j) -1]  \bigg\}
\label{18aa}
\end{eqnarray}
where $\epsilon = t/M$.
To evaluate the expression involving 
$N f[\{ \hat {\vec a}^\dagger \}, \{ \hat {\vec a} \}]$,
we use normal ordering.
We define 
\begin{eqnarray}
N f[\{ \hat {\vec a}^\dagger \}, \{ \hat {\vec a} \}]
=
N :f[\{ \hat {\vec a}^\dagger \}, \{ \hat {\vec a} \}] :
+ \Delta f[\{ \hat {\vec a}^\dagger \}, \{ \hat {\vec a} \}] 
\label{18a}
\end{eqnarray}
where the notation $: (\cdot) :$ means in the operator expression
for  $( \cdot )$, place all of the $\{ \hat {\vec a}^\dagger \}$
to the left of the $\{ \hat {\vec a} \}$.  The additional terms
that this operator commutation generates are collected in 
$\Delta f$.  We note that N $:f:$ is $O(N)$, whereas
$\Delta f$ is $O(1)$.
For example, for the quadratic replication rate 
$f(u) = \frac{\gamma}{2} u^2$, we find
\begin{eqnarray}
N f[\frac{1}{N} \sum_{j=1}^N  \hat {\vec a}^\dagger (j) 
\sigma_3  \hat {\vec a}(j)  ] &=&
\frac{N \gamma}{2} [ \frac{1}{N} \sum_{j=1}^N 
\hat {\vec a}^\dagger (j)
\sigma_3  \hat {\vec a}(j)]^2
\nonumber \\ &=&
\frac{\gamma}{2N} \sum_{i,j=1}^N 
\hat {\vec a}^\dagger (i) \sigma_3(i)  \hat {\vec a}(i)
\hat {\vec a}^\dagger (j) \sigma_3(j)  \hat {\vec a}(j)
\nonumber \\ &=&
\frac{\gamma}{2N} \sum_{i,j=1}^N 
\hat {\vec a}^\dagger (i) \sigma_3(i)  
[\hat {\vec a}^\dagger (j) \hat {\vec a}(i) + \delta_{ij}]
\sigma_3(j)  \hat {\vec a}(j)
\nonumber \\ &=&
\frac{N \gamma}{2} :[\frac{1}{N} \sum_{j=1}^N  \hat {\vec a}^\dagger (j)    
\sigma_3  \hat {\vec a}(j)]^2: + 
\frac{\gamma}{2N} \sum_{j=1}^N  
   \hat {\vec a}^\dagger (j) \cdot \hat {\vec a}(j)
\nonumber \\ &=&
N :f( \sum_{j=1}^N  \hat {\vec a}^\dagger (j) 
\sigma_3  \hat {\vec a}(j) /N ): + \frac{\gamma}{2}
\label{18b}
\end{eqnarray}
so that $\Delta f = \gamma / 2$ in this quadratic  case.
By induction, we can
show that the general form of the
commutation term is 
\begin{eqnarray}
\Delta f = \frac{1}{2} \frac{d^2 f(\xi) }{d \xi^2}
\label{18c}
\end{eqnarray}

In the continuous limit, the probability at time $t$ becomes
\begin{eqnarray}
P( \{ \vec n \}, t) &=&
\int [ {\cal D} \vec z^*
{\cal D} \vec z ]
\sum_{\{ \vec n_0 \} }
P(\{ \vec n_0 \})
\prod_{j =1}^N 
\vec n_j(t) \cdot \vec z_j(t)  ~
\vec z_j^*(0) \cdot \vec n_j(0)  ~
e^{ -S[\vec z^*, \vec z]}
\nonumber \\
\label{19}
\end{eqnarray}
where we have switched the subscripts and arguments of the variables and
where
\begin{eqnarray}
S[\vec z^*, \vec z] &=& \int_0^t dt'
\sum_{j=1}^N \vec z_j^*(t') [ 
   \partial_t - \mu \sigma_1(j) + \delta(t) ] \vec z_j(t')
+\mu N t
\nonumber \\ &&
- N \int_0^t dt'
 f\left[\sum_{j=1}^N  {\vec z}^*_j (t') 
\sigma_3  {\vec z}_j(t')
/N \right]
\nonumber \\ &&
- \int_0^t dt' \Delta f\left[ \{ {\vec z}^*(t')\}, \{ \vec z (t') \}\right]
\label{20}
\end{eqnarray}

At long times, 
we find that $e^{- \hat H t} \vert \{ {\bf n}_0 \} \rangle
\sim e^{f_m t} \vert {\bf n}^* \rangle$ by the Frobenius-Perrone
Theorem \cite{Bapat}, independent of
initial conditions, where $f_m$ is the unique largest
eigenvalue of $\hat H$, and $\vert {\bf n}^* \rangle$ is the
corresponding eigenvector.  To evaluate this eigenvalue,
 we consider the matrix trace.
We, furthermore, incorporate
the projection operator by 
the twisted boundary condition $\vec z_0(j) = e^{i \lambda_j} \vec z_M(j)$
arising from Eq.\ (\ref{17}).
We find
\begin{eqnarray}
{ Z} &=&  {\rm Tr} e^{- t \hat H} \hat P
\nonumber \\
&=&
\int_0^{2 \pi} \left[ \prod_{j=1}^N \frac{d \lambda_j}{2 \pi} e^{-i \lambda_j}
\right]
\lim_{M \to \infty} \int \left[ \prod_{k=1}^M \prod_{j=1}^N
\frac{d \vec z_k^*(j) d \vec z_k(j)}{\pi^2} \right]
e^{-S[\vec z^*, \vec z]}
\label{21}
\end{eqnarray}
where
\begin{eqnarray}
e^{-S[\vec z^*, \vec z]} =
\prod_{k=1}^M \langle
\{ \vec z_k \}
 \vert e^{-\epsilon \hat H }\vert
\{ \vec z_{k-1} \}
\rangle 
\label{22}
\end{eqnarray}
with boundary condition $\vec z_0(j) =e^{i \lambda_j} \vec z_M(j) $.
The action is Eq.\ (\ref{18aa}), without the initial 
$\vec z^*_0(j) \cdot \vec z_0 (j)$
term.
Since the replication rate depends only on the total magnetization,
the expression for $ Z$ can be simplified.  In particular,
we introduce $\xi_k = \frac{1} {N} \sum_{j=1}^N {\vec z}^*_{k} (j)
\sigma_3  {\vec z}_{k-1}(j)$ to find, as discussed in Appendix A,
that the partition function becomes
\begin{eqnarray}
 Z = \int [{\cal D} \bar \xi {\cal D} \xi ]
e^{-S[\bar \xi, \xi]}
\label{28}
\end{eqnarray}
where
\begin{eqnarray}
S[\bar \xi, \xi] = 
N \int_0^t dt' \left\{ - f[\xi(t')] + \bar \xi(t') \xi(t') + \mu - 
   \frac{1}{N} \Delta f  \right\}
- N \ln Q
\label{29}
\end{eqnarray}
and
\begin{eqnarray}
Q = {\rm Tr} \hat T e^{ \int_0^t  dt'
[ \mu \sigma_1 + \bar \xi(t') \sigma_3 ] }
\label{29a}
\end{eqnarray}

\subsection{The large $N$ limit of the parallel model is a saddle point}

The general expression of the parallel model partition function 
involves a functional integral.
Using that $N$ is large, this functional integral
can be evaluated by the saddle point method.  We impose
the saddle point condition to find
\begin{eqnarray}
\left. \frac{\delta S}{\delta \bar \xi} \right\vert_{\bar \xi_c, \xi_c}
&=& 0 =  N \xi_c - N
\frac{ {\rm Tr} \hat T \sigma_3 e^{ \int_0^t  dt'
[ \mu \sigma_1 + \bar \xi_c \sigma_3 ] } }
{ {\rm Tr} \hat T  e^{ \int_0^t  dt'
[ \mu \sigma_1 + \bar \xi_c \sigma_3 ] } }
\nonumber \\
\left. \frac{\delta S}{\delta  \xi}\right\vert_{\bar \xi_c, \xi_c} 
&=& 0 = -N f'(\xi_c) + N \bar \xi_c
\label{30}
\end{eqnarray}
Evaluating the traces, we find
\begin{eqnarray}
\bar \xi_c &=& f'(\xi_c)
\nonumber \\ 
\xi_c &=& \frac{\bar \xi_c}{ [\mu^2 + \bar \xi_c^2]^{1/2}}
 \tanh t [\mu^2 + \bar \xi_c^2]^{1/2}
\label{31}
\end{eqnarray}
For large $t$  we can solve Eq.\ (\ref{31}) for
$\bar \xi_c$ to find
\begin{eqnarray}
\bar \xi_c \sim \frac{\mu \xi_c}{  \sqrt{1  - \xi_c^2}}
\label{31a}
\end{eqnarray}
and evaluate Eq.\ (\ref{27}) to find
\begin{eqnarray}
\ln Q \sim t \sqrt{\mu^2 + \bar \xi_c^2}
= t \frac{\mu}{ \sqrt{1  - \xi_c^2} }
\label{31b}
\end{eqnarray}
Using Eqs.\ (\ref{31a}--\ref{31b}) in Eq.\ (\ref{29}),
we find that $\xi_c$ is the value which maximizes
\begin{eqnarray}
\frac{\ln Z}{t N} &=&
 f[\xi_c] - \frac{\mu \xi_c^2}{ \sqrt{1  - \xi_c^2} } - \mu
  + \frac{\mu}{ \sqrt{1  - \xi_c^2} }
   +\frac{\Delta f}{N} 
\nonumber \\ &=&
 f[\xi_c] 
   + \mu \sqrt{1  - \xi_c^2} 
- \mu
   +\frac{\Delta f}{N}
\label{32}
\end{eqnarray}
This expression is the saddle point evaluation of the
parallel model partition function.
It is valid
for arbitrary replication rate functions $f$.

As an example, we calculate the error threshold
for two different replication rate functions.
For our first example, we take the case of $f(1) = A$ and $f = 0$
otherwise.  This case leads to the phase transition at $A/ \mu = 1$.
For $A/\mu > 1$ a finite fraction $p_1$ of the population is at
$\xi_c=1$, whereas for $A / \mu < 1$, all of the population is
at $\xi_c = 0$.  
The fraction of the population at $\xi_c=1$
is determined by the implicit equation $p_1 f(1) + (1-p_1) f(\xi \ne 1) =
 \left. \ln Z / (t N) \right\vert_{\xi_c}$, which gives
$p_1 = 1 - \mu / A$.
  For our second example, we consider the
quadratic fitness $f(\xi) = k \xi^2 / 2$  \cite{bb97}.
  We find a phase
transition at $k / \mu = 1$, where the selected phase occurs for
$k / \mu > 1$ with an average magnetization given by
$\xi_c = \pm \sqrt{1 - (\mu/k)^2 }$.  The observable, surface
magnetization, $u_*$, is given by the implicit expression
$f(u_*) =  \left. \ln Z / (t N) \right\vert_{\xi_c}$ so that
$u_* = \pm (1 - \mu / k)$.

\subsection{$O(1/N)$ corrections to the parallel model}

\begin{figure}[t!]
\epsfig{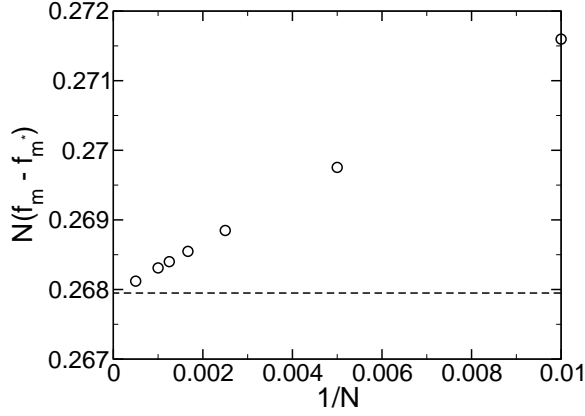}
\caption{The $O(1/N)$ shift in the free energy is
shown (circles).  Also shown is the prediction 
from Eq.\ (\ref{47}) (dashed line).  We use
$f(m) = k m^2/2$  with
$k/\mu = 2$ and $\mu = 1$.
\label{fig1}
}
\end{figure}

We now evaluate the fluctuation corrections to this result.
This procedure will determine the other $O(1/N)$
contributions to the mean replication rate per site,
$(\ln Z)/(t N)$.  We expand the action around the
saddle point limit
\begin{eqnarray}
S[ \bar \xi,  \xi ] &=& S(\bar \xi_c, \xi_c) +
\frac{1}{2} \sum_{k,l=1}^M  \bigg[
\left. \frac{\partial^2 S}{\partial \xi_k \partial \xi_{l}}
 \right\vert_{\bar \xi_c, \xi_c}
\delta \xi_k \delta \xi_{l}
+ 2 \left. \frac{\partial^2 S}{\partial \xi_k \partial \bar \xi_{l}} 
 \right\vert_{\bar \xi_c, \xi_c}
\delta \xi_k \delta \bar \xi_{l}
\nonumber \\ &&
+ \left. \frac{\partial^2 S}{\partial \bar \xi_k \partial \bar \xi_{l}} 
 \right\vert_{\bar \xi_c, \xi_c}
\delta \bar \xi_k \delta \bar \xi_{l}
\bigg]
\label{33}
\end{eqnarray}
We find
\begin{eqnarray}
\left. \frac{\partial^2 S}{\partial \xi_k \partial \xi_{l}} 
 \right\vert_{\bar \xi_c, \xi_c}
 &=&
-N \epsilon f''(\xi_c) \delta_{k l}
\nonumber \\
\left. \frac{\partial^2 S}{\partial \xi_k \partial \bar \xi_{l}}  
 \right\vert_{\bar \xi_c, \xi_c}
 &=&
 \epsilon N \delta_{k l}
\nonumber \\
\left. \frac{\partial^2 S}{\partial \bar \xi_k \partial \bar \xi_{l}} 
 \right\vert_{\bar \xi_c, \xi_c}
 &=&
- N \epsilon^2
 \frac{
 {\rm Tr}
  \sigma_3
e^{\epsilon \vert l-k \vert (\mu \sigma_1 + \bar \xi_c \sigma_3)} 
  \sigma_3
e^{\epsilon (M-\vert l-k \vert)(\mu \sigma_1 + \bar \xi_c \sigma_3)} 
}
{
 {\rm Tr} e^{\epsilon M(\mu \sigma_1 + \bar \xi_c \sigma_3)} 
}
\nonumber \\ && +
N \epsilon^2 \left(
 \frac{
 {\rm Tr}  \sigma_3
e^{\epsilon M (\mu \sigma_1 + \bar \xi_c \sigma_3)} 
}
{
 {\rm Tr} e^{\epsilon M(\mu \sigma_1 + \bar \xi_c \sigma_3)} 
}
\right)^2
\label{34}
\end{eqnarray}
These terms, and the matrix trace, are evaluated in Appendix B.
Solving the result of Eq.\ (\ref{45}) for 
$\bar \xi_c$ in terms of $\xi_c$, we find
\begin{eqnarray}
\frac{\ln Z}{t N} &=&
 f[\xi_c] 
   + \mu \sqrt{1  - \xi_c^2} 
- \mu
   +\frac{\Delta f}{N}
- \frac{ f''(\xi_c)}{2N} 
\nonumber \\ && +
\frac{1}{N} \frac{\mu}{ \sqrt{ 1 - \xi_c^2}}
\left[1 - \left[1 -  f''(\xi_c) (1-\xi_c^2)^{3/2} / \mu\right]^{1/2} \right]
\label{46}
\end{eqnarray}
Using Eq.\ (\ref{18c}) we find
\begin{eqnarray}
\frac{\ln Z}{t N} &=&
 f[\xi_c] 
   + \mu \sqrt{1  - \xi_c^2} 
- \mu
\nonumber \\ && +
\frac{1}{N} \frac{\mu}{ \sqrt{ 1 - \xi_c^2}}
\left[1 - \left[1 -  f''(\xi_c) (1-\xi_c^2)^{3/2} / \mu\right]^{1/2} \right]
\label{47}
\end{eqnarray}
This is the expression of the parallel model partition function
accurate to $O(1/N^2)$.  
The expression is accurate
for arbitrary smooth replication rate functions $f$.
Shown in Figure \ref{fig1} is the comparison between
this analytical result and a numerical calculation
following the algorithm in \cite{bb97}.

\section{Spin coherent state representation of the Eigen model}

\subsection{The Eigen model}

In the Eigen model, the probability distribution of viruses in the
space of all possible viral genomes is considered, as in
the parallel model.  However, when a virus
replication event occurs, the virus copies its genome, making mutations
at a rate of $1-q$ per base during the replication.  The probability 
distribution in genome space satisfies
\begin{eqnarray}
\frac{d p_i}{dt}=
\sum_{j=1}^{2^N}\left[B_{ij}r_j-\delta_{ij}D_j \right] p_j-p_i \left[
\sum_{j=1}^{2^N}(r_{j}-D_j)p_j \right] \ .
\label{48a} 
\end{eqnarray}
Here the  transition rates are given by
$B_{ij}=q^{N-d(i,j)}(1-q)^{d(i,j)}$.
We define the parameter $\mu = N(1-q)/q$ to characterize
the per genome replication rate.  We take $\mu = O(1)$.
We define $q^N = e^{- \mu_1}$ and note that in the
large $N$ limit, $\mu_1 \to \mu$.
As with the parallel model, the non-linear terms simply enforce 
conservation of probability, and it suffices to consider the
linear terms only
\begin{eqnarray}
\frac{d q_i}{dt}=
\sum_{j=1}^{2^N}\left[B_{ij}r_j-\delta_{ij}D_i \right] q_j
\label{48} 
\end{eqnarray}
As with the replication rate, the degradation rate is
defined by $D_i = N d(u)$, where $u = (1/N) \sum_n s^i_n$.

\subsection{The Eigen model in operator form}

Using the creation and annihilation operator formalism, the
dynamics can be again written in the form of Eq.\ (\ref{9}), with
\begin{eqnarray}
-\hat H &=& 
\prod_{j=1}^N [q T_0(j) + (1-q) T_1(j) ]
N f\left[\sum_{k=1}^N T_3(k)/N \right] 
- N  d \left[\sum_{j=1}^N T_3(j)/N \right] 
\nonumber \\ &=&
N e^{-\mu_1} \prod_{j=1}^N [1 + \frac{\mu}{N} T_1(j) ]
f\left[\sum_{k=1}^N T_3(k)/N \right] 
- N  d \left[\sum_{j=1}^N T_3(j)/N \right] 
\nonumber \\ &\sim&
N e^{-\mu_1}
 e^{\sum_{j=1}^N \mu T_1(j)/N}
f\left[\sum_{k=1}^N T_3(k)/N \right] 
- N  d \left[\sum_{j=1}^N T_3(j)/N \right] 
\label{49}
\end{eqnarray}
where the last expression is valid for large $N$.  Corrections
to this expression will come from $\mu_1 \ne \mu$, the exponential
not being exactly equal to the product, as well as normal ordering
terms.  We will address these corrections later.

\subsection{The field theoretic representation of the Eigen model}

We introduce the Schwinger spin coherent states.
We consider the normal ordered form of the Hamiltonian,
and first consider the expression $:\hat H:$.  We will
consider the commutator terms later.
We find $P(\{ \vec n \},t)$ can be expressed as in Eq.\ (\ref{18})
and the partition function $ Z$ can be expressed as in
Eq.\ (\ref{21}) with 
\begin{eqnarray}
S[\vec z^*, \vec z] &=&
\prod_{k=1}^M \langle
\{ \vec z_k \}
 \vert e^{-\epsilon : \hat H : }\vert
\{ \vec z_{k-1} \}
\rangle 
\nonumber \\ &=&
\vec z^*_0(j) \cdot \vec z_0 (j)
+ \sum_{k =1}^M
{\vec z}_k^* (j) \cdot [\vec z_k(j)- \vec z_{k-1}(j)]
\nonumber \\ &&
+ \epsilon N \sum_{k =1}^M \bigg\{
 e^{-\mu_1} e^{\mu \sum_{j=1}^N {\vec z}_k^* (j)  \sigma_1
 \vec z_{k-1}(j) / N}
f\left[
\sum_{j=1}^N  {\vec z}^*_{k} (j) 
\sigma_3  {\vec z}_{k-1}(j) /N 
\right]
\nonumber \\ && 
-  d\left[
\sum_{j=1}^N  {\vec z}^*_{k} (j)
\sigma_3  {\vec z}_{k-1}(j) /N  
\right] \bigg\}
\label{50}
\end{eqnarray}
For the partition function case, we have
the boundary condition $\vec z_0(j) =e^{i \lambda_j} \vec z_M(j) $.
We introduce 
$\xi_k = \frac{1} {N} \sum_{j=1}^N {\vec z}^*_{k} (j)
\sigma_3  {\vec z}_{k-1}(j)$ 
and
$\eta_k = \frac{1} {N} \sum_{j=1}^N {\vec z}^*_{k} (j)
\sigma_1  {\vec z}_{k-1}(j)$ 
to find, as discussed in Appendix C,
that
the partition function becomes
\begin{eqnarray}
Z = \int 
[{\cal D} \bar \xi {\cal D} \xi ]
[{\cal D} \bar \eta {\cal D} \eta ]
e^{-S[\bar \xi, \xi, \bar \eta, \eta]}
\label{54}
\end{eqnarray}
where
\begin{eqnarray}
S[\bar \xi, \xi, \bar \eta, \eta] &= &
N \int_0^t dt' \left\{ 
 - e^{-\mu_1} e^{ \mu \eta(t')}
f[\xi(t')]  + d[\xi(t')] +  \bar \xi(t') \xi(t')
+ \bar \eta(t') \eta(t')
 \right\}
\nonumber \\ &&
- N \ln Q
\label{55}
\end{eqnarray}

\subsection{The large $N$ limit of the Eigen model is a saddle point}

The partition function of the Eigen model is represented
as a functional integral.  In the limit of large $N$, this
integral can be evaluated by the saddle point method.
The saddle point conditions are
\begin{eqnarray}
\left. \frac{\delta S}{\delta \bar \xi}
 \right\vert_{\bar \xi_c, \xi_c, \bar \eta_c, \eta_c}
&=& 0 =  N \xi_c - N
\frac{ {\rm Tr} \hat T \sigma_3 e^{ \int_0^t  dt'
[ \bar \eta_c \sigma_1 + \bar \xi_c \sigma_3 ] } }
{ {\rm Tr} \hat T  e^{ \int_0^t  dt'
[ \bar \eta_c \sigma_1 + \bar \xi_c \sigma_3 ] } }
\nonumber \\
\left. \frac{\delta S}{\delta  \xi}
 \right\vert_{\bar \xi_c, \xi_c, \bar \eta_c, \eta_c}
&=& 0 = -N e^{-\mu_1} e^{ \mu \eta_c} f'(\xi_c)
+ N d'(\xi_c) + N \bar \xi_c
\nonumber \\
\left. \frac{\delta S}{\delta \bar \eta}
 \right\vert_{\bar \xi_c, \xi_c,\bar \eta_c, \eta_c}
&=& 0 =  N \eta_c - N
\frac{ {\rm Tr} \hat T \sigma_1 e^{ \int_0^t  dt'
[ \bar \eta_c \sigma_1 + \bar \xi_c \sigma_3 ] } }
{ {\rm Tr} \hat T  e^{ \int_0^t  dt'
[ \bar \eta_c \sigma_1 + \bar \xi_c \sigma_3 ] } }
\nonumber \\
\left. \frac{\delta S}{\delta  \eta} 
\right\vert_{\bar \xi_c, \xi_c, \bar \eta_c, \eta_c}
&=& 0 = -N \mu e^{-\mu_1} e^{ \mu \eta_c} f(\xi_c) + N \bar \eta_c
\label{56}
\end{eqnarray}
Evaluating the traces, we find
\begin{eqnarray}
\bar \xi_c &=& e^{-\mu_1} e^{ \mu \eta_c} f'(\xi_c) - d'(\xi_c)
\nonumber \\ 
\xi_c &=& \frac{\bar \xi_c}{ [\bar \eta_c^2 + \bar \xi_c^2]^{1/2}}
 \tanh t [\bar \eta_c^2 + \bar \xi_c^2]^{1/2}
\nonumber \\
\bar \eta_c &=& \mu e^{-\mu_1} e^{ \mu \eta_c} f(\xi_c)
\nonumber \\
\eta_c &=&
 \frac{\bar \eta_c}{ [\bar \eta_c^2 + \bar \xi_c^2]^{1/2}}
 \tanh t [\bar \eta_c^2 + \bar \xi_c^2]^{1/2}
\label{57}
\end{eqnarray}
At long time, we find 
\begin{eqnarray}
\eta_c &=& \sqrt{1 - \xi_c^2}
\nonumber \\
\bar \eta_c \eta_c + \bar \xi_c \xi_c &=& \sqrt{\bar \eta_c^2 + \bar \xi_c^2}
\nonumber \\
\ln Q &=& t \sqrt{\bar \eta_c^2 + \bar \xi_c^2}
\label{57a}
\end{eqnarray}
from the second and fourth lines of Eq.\ (\ref{57})
and from Eq.\ (\ref{53}).
Using Eq. (\ref{57a}) in Eq.\ (\ref{55}),
we find that $\xi_c$ is the value which maximizes
\begin{eqnarray}
\frac{\ln Z}{t N} &=&
e^{-\mu_1} e^{ \mu \sqrt{1 - \xi_c^2}} f(\xi_c)
- d(\xi_c)
\label{58}
\end{eqnarray}
This is the saddle point expression for the Eigen model
partition function.  
It is valid
for arbitrary replication rate functions $f$ and
degradation functions $d$.

As an example, we calculate the error threshold
for two different  replication rate functions.
For our first example, we take the case of $f(1) = A$ and $f = 1$
otherwise.  This case leads to the phase transition at $A e^{- \mu} = 1$
\cite{Deem2006}.
For $A e^{- \mu} >1 $ a finite fraction $p_1$ of the population is at
$\xi_c=1$, whereas for $A e^{- \mu}  < 1$, all of the population is
at $\xi_c = 0$.  
The fraction of the population at $\xi_c=1$
is determined by the implicit equation $p_1 f(1) + (1-p_1) f(\xi \ne 1) =
 \left. \ln Z / (t N) \right\vert_{\xi_c}$, which gives
$p_1 = (A e^{- \mu} - 1)/(A-1)$.
For our second example, we consider the
quadratic fitness $f(\xi) = 1 + k \xi^2 / 2$
\cite{Deem2006}.
  We find a phase
transition at $k / \mu = 1$, where the selected phase occurs for
$k / \mu > 1$ with magnetization
$\xi_c = \pm \sqrt 2 [  \sqrt{1 + \mu^2 (1 + 2/k )} 
-  1 - \mu^2/k ]^{1/2} / \mu$.  The observable, surface
magnetization, $u_*$, is given by the implicit expression
$f(u_*) =  \left. \ln Z / (t N) \right\vert_{\xi_c}$, which for the
Eigen model is a non-linear, transcendental equation even for
the quadratic fitness case.

\subsection{$O(1/N)$ corrections to the Eigen model}

There are $O(1/N)$ corrections to Eq.\ (\ref{58}).  The first
comes from 
\begin{eqnarray}
e^{-\mu_1} = e^{-\mu} \left(1 +\frac{\mu^2}{2 N} \right)
\label{59}
\end{eqnarray}
The second comes from the normal ordering of $f$ and $d$:
\begin{eqnarray}
\Delta f &=& \frac{1}{2} \frac{d^2 f(\xi)  }{d \xi^2}
\nonumber \\
\Delta d &=& \frac{1}{2}\frac{d^2 d(\xi)  }{d \xi^2}
\label{60}
\end{eqnarray}
  The third term comes from the approximation
made in the last line of Eq.\ (\ref{49}).
\begin{eqnarray}
\prod_{j=1}^N [1 + \frac{\mu}{N} T_1(j) ]
 &=& \prod_{j=1}^N  
:e^{ \mu T_1(j)/N}:
\nonumber \\ &=&
:e^{\sum_{j=1}^N \mu T_1(j)/N }: 
\nonumber \\ & \to &
e^{\mu \eta}
\label{62}
\end{eqnarray}
since $:T_\alpha(j)^n: ~ = 0$ for $n >1$ due to
constraint (\ref{3a}).
The fourth term comes from normal ordering $T_1(j)$ 
in $\exp[ \sum_{j=1}^N \mu T_1(j)/N]$ and 
$T_3(j)$ in $f[\sum_{j=1}^N T_3(j)/N]$ in
$f$:
\begin{eqnarray}
&&
:e^{-\mu} e^{\sum_{j=1}^N \mu T_1(j)/N }:
f\left[\sum_{i=1}^N T_3(i)/N \right] 
\nonumber \\ &=&
:e^{-\mu} e^{\sum_{j=1}^N \mu T_1(j)/N }
f\left[\sum_{i=1}^N T_3(i)/N \right] :
\nonumber \\ &&
+ \left. \frac{d e^{-\mu} e^{\mu \eta}}
     {d \eta} \right\vert_{\eta =\sum_{j=1}^N  T_1(j)/N}
 \left. \frac{d f(\xi)}
     {d \xi} \right\vert_{\xi =\sum_{j=1}^N T_3(j)/N}
\nonumber \\ && \times
\frac{1}{N} \sum_{j=1}^N  [T_1(j) T_3(j) - : T_1(j) T_3(j): ]/N
\nonumber \\ &=&
:e^{-\mu} e^{\sum_{j=1}^N \mu T_1(j)/N }
f\left[\sum_{i=1}^N T_3(i)/N \right] :
\nonumber \\ &&
+ \left. \frac{d e^{-\mu} e^{\mu \eta}}
     {d \eta} \right\vert_{\eta =\sum_{j=1}^N  T_1(j)/N}
 \left. \frac{d f(\xi)}
     {d \xi} \right\vert_{\xi =\sum_{j=1}^N T_3(j)/N}
\frac{1}{N} \sum_{j=1}^N  T_2(j) /N
\nonumber \\ &\to&
e^{-\mu} e^{\mu \eta_k} f(\xi_k)
+
\frac{\mu}{N} e^{-\mu} e^{\mu \eta_k} f'(\xi_k) \kappa_k
\label{61}
\end{eqnarray}
where $\kappa_k = 
 \frac{1} {N} \sum_{j=1}^N {\vec z}^*_{k} (j)
\sigma_2  {\vec z}_{k-1}(j)$ .
Introducing this new field, we find $\bar \kappa_c = 0$ at the
saddle point.  Moreover, we find
\begin{eqnarray}
\kappa_c
= \frac{ {\rm Tr} \hat T \sigma_2 e^{ \int_0^t  dt'
[ \bar \eta \sigma_1 + \bar \xi_c \sigma_3 ] } }
{ {\rm Tr} \hat T  e^{ \int_0^t  dt'
[ \bar \eta \sigma_1 + \bar \xi_c \sigma_3 ] } } = 0
\label{61a}
\end{eqnarray}
The trace evaluates as $\langle \sigma_2 \rangle=0$.
Thus this fourth, commutator term vanishes.

\begin{figure}[t!]
\epsfig{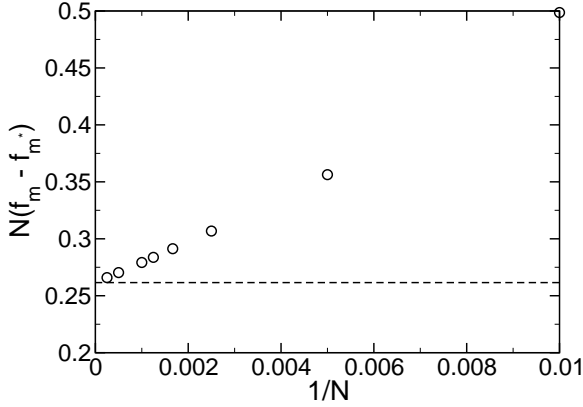}
\caption{The $O(1/N)$ shift in the free energy is
shown (circles).  Also shown is the prediction 
from Eq.\ (\ref{84}) (dashed line).  We use
$f(m) = k m^2/2 + 1$ and $d(m) = 0$ with
$k/\mu = 2$ and $\mu = 5$.
\label{fig2}
}
\end{figure}

There are also fluctuation corrections to Eq.\ (\ref{58}).
These are discussed in Appendix D.
Using  the results of Eqs.\ (\ref{57}), (\ref{57a}), and (\ref{83})
we find
\begin{eqnarray}
\frac{\ln Z}{t N} 
&=&
e^{-\mu+ \mu \sqrt{1 - \xi_c^2}} f(\xi_c) - d(\xi_c)
+ \frac{1}{N} 
b' \left [
1 - \left[ 1 - c'/{b'}    \right]^{1/2}
\right]
\nonumber \\
\label{84}
\end{eqnarray}
where the constants $a'$, $b'$, and $c'$ are defined
in Eq.\ (\ref{81cc}).
This is the expression of the Eigen model partition function
accurate to $O(1/N^2)$.
The expression is accurate
for arbitrary smooth replication rate functions $f$ and
degradation functions $d$.
Shown in Figure \ref{fig2} is the comparison between
this analytical result and a numerical calculation
following the algorithm in \cite{ei89}.

\section{The full probability distributions for continuous-time
quasispecies theory}

In this section we derive the field theoretic
expressions for the full probability distribution 
functions of the parallel and Eigen quasispecies
theories.  By use of the coherent states formalism, we
are able to derive the distribution for arbitrary initial
and final conditions.  In the long time limit, the initial
condition will not matter, as the system will reach
a steady state.  The final condition matters, though, because
the weight assigned by the field theory
to a given final condition is exactly
equal to the probability that a given surface magnetization
occurs in the population of viruses.

\subsection{Field theoretic representation of the 
full probability distribution of the parallel model}

To obtain the full probability distribution $P(\{ \vec n \}, t)$,
rather than simply the largest Frobenius-Perrone eigenvalue, for the 
parallel model from Eq.\ (\ref{18})
we add a term 
\begin{eqnarray}
\delta S = 
- \sum_{j=1}^N [\vec J^*(j) \cdot \vec z_M(j) +
              \vec J_0(j) \cdot \vec z_0^*(j)]
\label{63}
\end{eqnarray}
to the action.
We find
\begin{eqnarray}
P(\{ \vec n \}, t) = 
\sum_{\{ \vec n_0 \} }
P(\{ \vec n_0 \})
\left. \prod_{j=1}^N
\frac{\partial }{\partial J^*_{\alpha(j)}(j)}
\frac{\partial }{\partial {J_0}_{\alpha_0(j)}(j)}
\right\vert_{J^* = J_0 = 0}
{ Z}
\label{64}
\end{eqnarray}
Here $\alpha(j) = 2-n(j)$ is the value of the spin at the
final time, and $\alpha_0(j) = 2-n_0(j)$ is the value of the
spin at the initial time.  Evaluation of this expression is
carried out in Appendix E.  Here we use the result of
Appendix E for a
couple different types of initial conditions.
We define $Q_{ij}$ to be the matrix element of the
matrix $ \hat T e^{ \int_0^t  dt'
[ \mu \sigma_1 + \bar \xi(t') \sigma_3 ] }$.

If, for example, we say that at $t=0$, the spins
are distributed randomly, but with a given
initial surface magnetization, $u_0$, then
we take the terms in the multinomial expansion of
$(Q_{11} + 
Q_{12} + 
Q_{21} + 
Q_{22} )^N$ that satisfy the initial
and final conditions:
\begin{eqnarray}
P(u, t \vert u_0) 
&=& \lim_{M \to \infty}
 \int \left[ \prod_{k=1}^M  \frac{i \epsilon N d \bar \xi_k d \xi_k }{2 \pi}
\right]
e^{\epsilon \sum_{k=1}^M 
[ N f(\xi_k) - N \bar \xi_k \xi_k - N \mu + \Delta f] }
\nonumber \\ &&
\times
\sum_{j_1 = \max[0, \frac{N(u_0 + u)}{2}]}^{\min[N, 
       \frac{N(1+u)}{2}, \frac{N(1+u_0)}{2}]}
\frac{N!}{j_1! j_2! j_3! j_4!}
Q_{11}^{j_1}
Q_{12}^{j_2}
Q_{21}^{j_3}
Q_{22}^{j_4}
\label{100}
\end{eqnarray}
where $j_1 + j_2 = N(1+u_0)/2$,
$j_1 + j_3 = N(1+u)/2$, and
$j_1 +j_2 + j_3 +j_4 = N$.

Alternatively, if we take
an initial condition with spin up and down
equally likely ($\langle u_0 \rangle = 0$) and define
$Q_+ = Q_{11} + Q_{21}$
and
$Q_- = Q_{12} + Q_{22}$ we find
\begin{eqnarray}
P(u, t) 
&=& \lim_{M \to \infty}
 \int \left[ \prod_{k=1}^M  \frac{i \epsilon N d \bar \xi_k d \xi_k }{2 \pi}
\right]
e^{\epsilon \sum_{k=1}^M 
[ N f(\xi_k) - N \bar \xi_k \xi_k - N \mu + \Delta f] }
\nonumber \\ &&
\times
\left(
\begin{array}{c}
N \\ N (1+u)/2
\end{array}
\right)
Q_+[\bar \xi]^{N \frac{1+u}{2} }
Q_-[\bar \xi]^{N \frac{1-u}{2} }
\label{200}
\end{eqnarray}

\subsection{The large $N$ limit of the full probability distribution for
the parallel model}

Since the full probability distribution is also expressed
as a functional integral, it can be evaluated by a saddle point 
in the large $N$ limit.
In the saddle point limit, this equation becomes
\begin{eqnarray}
\frac{\ln P(u, t) }{N} &=&
\epsilon \sum_{k=1}^M
[  f(\xi_k) -  \bar \xi_k \xi_k -  \mu ]
- \frac{1+u}{2} \ln \frac{1+u}{2} 
- \frac{1-u}{2} \ln \frac{1-u}{2} 
\nonumber \\ &&
+ \frac{1+u}{2} \ln Q_+
+ \frac{1-u}{2} \ln Q_-
\label{210}
\end{eqnarray}
with
\begin{eqnarray}
\bar \xi_k &= & f'(\xi_k)
\nonumber \\ 
\xi_k &= &
 \frac{1+u}{2}
\frac{
\prod_{l=1}^{k-1} [I + \epsilon \mu \sigma_1 + \epsilon \bar \xi_l \sigma_3 ]
\sigma_3
\prod_{l=k+1}^{M} [I + \epsilon \mu \sigma_1 + \epsilon \bar \xi_l \sigma_3 
]_{11 + 21}
}{
\prod_{l=1}^{M} [I + \epsilon \mu \sigma_1 + \epsilon \bar \xi_l \sigma_3 
]_{11 + 21}
}
\nonumber \\ && 
+ \frac{1-u}{2}
\frac{
\prod_{l=1}^{k-1} [I + \epsilon \mu \sigma_1 + \epsilon \bar \xi_l \sigma_3 ]
\sigma_3
\prod_{l=k+1}^{M} [I + \epsilon \mu \sigma_1 + \epsilon \bar \xi_l \sigma_3 
]_{12 + 22}
}{
\prod_{l=1}^{M} [I + \epsilon \mu \sigma_1 + \epsilon \bar \xi_l \sigma_3 
]_{12 + 22}
}
\nonumber \\ &\equiv& 
 \frac{1+u}{2} \langle \sigma_3(k) \rangle_+
+ \frac{1-u}{2} \langle \sigma_3(k) \rangle_-
\label{211}
\end{eqnarray}
In Appendix F we evaluate these expressions where the
probability distribution is large, in the Gaussian central region.

\subsection{The parallel model distribution function in the
Gaussian central region}

Adding the terms from Appendix F together, 
we find that the probability distribution becomes
Gaussian in the central region.  That is Eq.\ (\ref{210}) becomes
\begin{eqnarray}
\frac{\ln P(u, t) }{N} &=&
({\rm const}) - \frac{1}{2} \delta u^2 \left( \frac{f'(u_*)}{2 \mu u_*}
\right)
\label{218}
\end{eqnarray}
We, therefore, conclude that
\begin{eqnarray}
\left\langle \left ( u - \langle u \rangle \right)^2 \right\rangle =
 \frac{2 \mu u_*}{N f'(u_*)}
\label{219}
\end{eqnarray}

The fluctuation correction to the fitness is given by
\begin{eqnarray}
\langle f(u  )  \rangle &= &
f( u_*  )  + 
f'(u_*)
\langle u - u_* \rangle 
+ \frac{1}{2} f''(u_*)
\langle (\delta u)^2 \rangle + O\left( \frac{1}{N^2}\right)
\label{202}
\end{eqnarray}
From expressions (\ref{47}) and (\ref{219}), we conclude
that there is also a shift of the average magnetization
for finite $N$:
\begin{eqnarray}
\langle u - u_* \rangle  &=&
\frac{1}{N f'(u_*)}\bigg[
 \frac{\mu}{ \sqrt{ 1 - \xi_c^2}}
\left[1 - \left[1 -  f''(\xi_c) (1-\xi_c^2)^{3/2} / \mu\right]^{1/2} \right]
\nonumber \\ &&
- \frac{\mu u_* f''(u_*)}{f'(u_*)}
\bigg]
\label{202a}
\end{eqnarray}

\subsection{An alternative derivation of the fluctuation corrections to the
parallel model distribution function}

As an alternative, we may compute averages of the magnetization
from the full functional integral expression for the
partition function.
Computing the surface magnetization From Eq.\ (\ref{200}), we find
\begin{eqnarray}
\langle u \rangle(t) 
&=& \frac{ \sum_{u'} u' P(u', t )}{\sum_{u'} P(u',t )}
\nonumber \\ &=&
\lim_{M \to \infty}
 \int \left[ \prod_{k=1}^M  \frac{i \epsilon N d \bar \xi_k d \xi_k }{2 \pi}
\right]
e^{\epsilon \sum_{k=1}^M 
[ N f(\xi_k) - N \bar \xi_k \xi_k - N \mu + \Delta f] }
\nonumber \\ && \times
(Q_+[\bar \xi] + Q_-[\bar \xi])^N
 \frac{Q_+[ \bar \xi] - Q_-[ \bar \xi]}
{Q_+[ \bar \xi] + Q_-[ \bar \xi]} 
\nonumber \\ && 
\bigg/
\lim_{M \to \infty}
 \int \left[ \prod_{k=1}^M  \frac{i \epsilon N d \bar \xi_k d \xi_k }{2 \pi}
\right]
e^{\epsilon \sum_{k=1}^M 
[ N f(\xi_k) - N \bar \xi_k \xi_k - N \mu + \Delta f] }
\nonumber \\ && \times
(Q_+[\bar \xi] + Q_-[\bar \xi])^N
\label{69}
\end{eqnarray}
This expression, however, is not easily calculated in the saddle point limit.

Computing the fluctuation, we find
\begin{eqnarray}
\langle u^2 \rangle(t) 
&=& \frac{ \sum_{u'} {u'}^2 P(u', t )}{\sum_{u'} P(u',t )}
\nonumber \\ &=&
\lim_{M \to \infty}
 \int \left[ \prod_{k=1}^M  \frac{i \epsilon N d \bar \xi_k d \xi_k }{2 \pi}
\right]
e^{\epsilon \sum_{k=1}^M 
[ N f(\xi_k) - N \bar \xi_k \xi_k - N \mu + \Delta f] }
\nonumber \\ && \times
(Q_+[\bar \xi] + Q_-[\bar \xi])^N
\left[
 \frac{(Q_+[ \bar \xi] - Q_-[ \bar \xi])^2 + 4 Q_+[ \bar \xi] Q_-[ \bar \xi]/N}
{(Q_+[ \bar \xi] + Q_-[ \bar \xi])^2} 
\right]
\nonumber \\ && 
\bigg/
\lim_{M \to \infty}
 \int \left[ \prod_{k=1}^M  \frac{i \epsilon N d \bar \xi_k d \xi_k }{2 \pi}
\right]
e^{\epsilon \sum_{k=1}^M 
[ N f(\xi_k) - N \bar \xi_k \xi_k - N \mu + \Delta f] }
\nonumber \\ && \times
(Q_+[\bar \xi] + Q_-[\bar \xi])^N
\label{205}
\end{eqnarray}
This equation implies
\begin{eqnarray}
\left\langle \left ( u - \langle u \rangle \right)^2 \right\rangle =
\left\langle 
\left( \delta \frac{Q_+ - Q_-}{Q_+ Q_-} \right)^2
 \right\rangle
+ \frac{1 - u_*^2}{N}
+ O\left( \frac{1}{N^2} \right)
\label{220}
\end{eqnarray}
The term $(1-u_*^2)/N$ is the variance of
the spin at a given site $1 \le i \le N$.  The term
$\langle (\delta \Delta Q/Q)^2 \rangle$, therefore,
is $N-1$ times
the $O(1/N)$ correlations between spins at
any two different sites $1 \le i  < j \le N$.
Base compositions at different sites, while equivalent,
are not uncorrelated.  There is a $O(1/N)$ correlation
between base compositions at different sites.

\begin{figure}[t!]
\epsfig{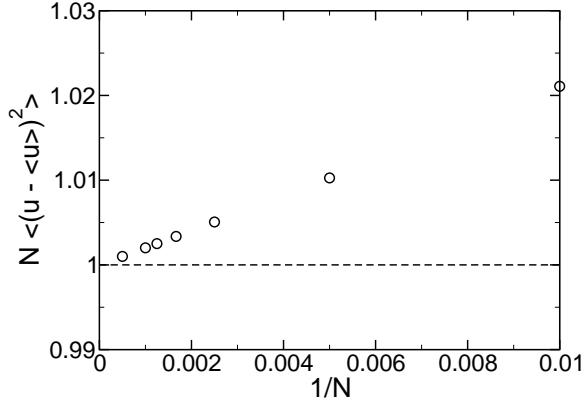}
\caption{The $O(1/N)$ shift in the variance of the
surface magnetization is
shown (circles).  Also shown is the prediction 
from Eq.\ (\ref{219}) (dashed line).  We use
$f(m) = k m^2/2 $ with
$k/\mu = 2$ and $\mu = 1$.
\label{fig3}
}
\end{figure}

The variance can be written as
\begin{eqnarray}
\left\langle \left ( u - \langle u \rangle \right)^2 \right\rangle =
\frac{
\int d (\delta u) e^{- N f'(u_*) \delta u^2 / (4 u_*)}
 e^{\delta S}
\left( \delta \frac{Q_+ - Q_-}{Q_+ Q_-} \right)^2
}
{
\int d (\delta u) e^{- N f'(u_*) \delta u^2 / (4 u_*)}
 e^{\delta S}
}
\label{221}
\end{eqnarray}
where
\begin{eqnarray}
\delta S/N &=& \ln (Q_+ + Q_-) 
+ \frac{1+u}{2} \ln \frac{1+u}{2} 
+ \frac{1-u}{2} \ln \frac{1-u}{2} 
\nonumber \\ && 
- \frac{1+u}{2} \ln Q_+
- \frac{1-u}{2} \ln Q_-
\label{222}
\end{eqnarray}
We note that
\begin{eqnarray}
\frac{\delta S}{N} =
 \frac{1}{2} \frac{1 - u_*^2}{4 u_*^2} f'(u_*)^2 \delta u^2 + O(\delta u^3)
\label{223}
\end{eqnarray}
Using Eqs.\ (\ref{216}), (\ref{217}), (\ref{221}), and
(\ref{223}) we find
exactly Eq.\ (\ref{219}), which 
shows the consistency of the two calculations.
This second calculation, however, makes explicit the
contributions of spin-spin correlations to the fluctuation
of the population in genome space.
Shown in Figure \ref{fig3} is the comparison between
the analytical result of  Eq.\ (\ref{219}) and a numerical calculation
following the algorithm in \cite{bb97}.

\subsection{Correlation functions of the parallel model field theory}

Since the theory
is Gaussian in the saddle point limit, we can evaluate correlations
from the inverse of the Hessian in
Eq.\ (\ref{34}).  
The inverse is given by
\begin{eqnarray}
\chi &=& 
\left(
\begin{array}{cc}
- N \epsilon f''(\xi_c) I & N \epsilon I
\\
\epsilon N I & - N \epsilon^2 \frac {\mu^2}{\mu^2 + \bar \xi_c^2} M
+ N \epsilon^2 I
\end{array}
\right)^{-1}
\label{204}
\end{eqnarray}
From Eq.\ (\ref{42a}) we find
\begin{eqnarray}
\hat \chi(k) &=& 
\frac{1}{N \epsilon} \left(
\begin{array}{cc}
- f''(\xi_c)  &  1
\\
1 & 
- \frac{4 \epsilon^2 \mu^2 }{ b}
\frac{1}{4 b^2 \epsilon^2 + k^2}
+ \epsilon
\end{array}
\right)^{-1}
\label{230}
\end{eqnarray}
Inverting this matrix, and inverse
Fourier transforming, we find
\begin{eqnarray}
\langle 
\delta \xi_k 
\delta \xi_l 
 \rangle &=&
-\frac{1}{N} \delta_{kl}
\nonumber \\ &&
+\frac{1}{N}
\frac{1-\xi_c^2}
{\sqrt{1 - (1-\xi_c^2)^{3/2} f''(\xi_c)/\mu}}
e^{- 2 \epsilon \vert k - l \vert \sqrt{b^2-f''(\xi_c) \mu^2/b} }
\nonumber \\
\langle 
\delta \bar \xi_k 
\delta \bar \xi_l 
 \rangle &=&
\frac{f''(\xi_c)}{N \epsilon} \delta_{kl}
- \frac{f''(\xi_c)^{2}}{N} \delta_{kl}
\nonumber \\ &&
+\frac{f''(\xi_c)^2}{N}
\frac{1-\xi_c^2}
{\sqrt{1 - (1-\xi_c^2)^{3/2} f''(\xi_c)/\mu}}
e^{- 2 \epsilon \vert k - l \vert \sqrt{b^2-f''(\xi_c) \mu^2/b} }
\nonumber \\
\nonumber \\
\langle 
\delta  \xi_k 
\delta \bar \xi_l 
 \rangle &=&
\frac{1}{N \epsilon} \delta_{kl}
- \frac{f''(\xi_c)}{N} \delta_{kl}
\nonumber \\ &&
+ \frac{f''(\xi_c)}{N}
\frac{1-\xi_c^2}
{\sqrt{1 - (1-\xi_c^2)^{3/2} f''(\xi_c)/\mu}}
e^{- 2 \epsilon \vert k - l \vert \sqrt{b^2-f''(\xi_c) \mu^2/b} }
\nonumber \\
\label{231}
\end{eqnarray}
for $k,l \gg 1$ and $k,l \ll M$,
where $b = \mu / \sqrt{1-\xi_c^2}$.

\subsection{Field theoretic representation of the full probability distribution of the Eigen model}

The full probability distribution for the Eigen model is
expressed as a functional integral in an analogous fashion
to the parallel model.
For the Eigen model we find 
\begin{eqnarray}
P(\{ \vec n \}, t) 
&=& \lim_{M \to \infty}
 \int \left[ \prod_{k=1}^M  
\frac{i \epsilon N d \bar \xi_k d \xi_k }{2 \pi}
\frac{i \epsilon N d \bar \eta_k  d \eta_k }{2 \pi}
\right]
\nonumber \\ &&\times
e^{\epsilon N \sum_{k=1}^M 
[ e^{-\mu + \mu \eta_k} f(\xi_k) - d(\xi_k)
- \bar \xi_k \xi_k 
- \bar \eta_k \eta_k
  }
\nonumber \\ &&
\times
\sum_{\{ \vec n_0 \} }
P(\{ \vec n_0 \})
\prod_{j=1}^N \{ [Q(j)]_{\alpha_0(j) \alpha(j)} \}
\label{240}
\end{eqnarray}
where
\begin{eqnarray}
Q(j) &=& 
\prod_{k=1}^M [I + \epsilon \bar \eta_k \sigma_1 +
 \epsilon \bar \xi_k \sigma_3 ]
\nonumber \\ &\sim&
\hat T e^{\epsilon \sum_{k=1}^M
[ \bar \eta_k \sigma_1 + \bar \xi_k \sigma_3 ] }
\nonumber \\ &=&
 \hat T e^{ \int_0^t  dt'
[ \bar \eta(t') \sigma_1 + \bar \xi(t') \sigma_3 ] }
\label{241}
\end{eqnarray}
Taking an initial condition with spin up and down
equally likely, we find
\begin{eqnarray}
P(u, t) 
&=& \lim_{M \to \infty}
 \int \left[ \prod_{k=1}^M  
\frac{i \epsilon N d \bar \xi_k d \xi_k }{2 \pi}
\frac{i \epsilon N d \bar \eta_k  d \eta_k }{2 \pi}
\right]
\nonumber \\  &&\times
e^{\epsilon N \sum_{k=1}^M 
[ e^{-\mu + \mu \eta_k} f(\xi_k) - d(\xi_k)
- \bar \xi_k \xi_k - \bar \eta_k \eta_k
  }
\nonumber \\ &&
\times
\left(
\begin{array}{c}
N \\ N (1+u)/2
\end{array}
\right)
Q_+[\bar \xi, \bar \eta]^{N \frac{1+u}{2} }
Q_-[\bar \xi, \bar \eta]^{N \frac{1-u}{2} }
\label{242}
\end{eqnarray}

\subsection{The large $N$ limit of the full probability distribution function
of the Eigen model}

Since the distribution function is expressed as a functional integral, it
can be evaluated by the saddle point method.
In the saddle point limit, the probability distribution
function becomes
\begin{eqnarray}
\frac{\ln P(u, t) }{N} &=&
\epsilon \sum_{k=1}^M
[ e^{-\mu + \mu \eta_k} f(\xi_k) - d(\xi_k)
- \bar \xi_k \xi_k - \bar \eta_k \eta_k ]
\nonumber \\ &&
- \frac{1+u}{2} \ln \frac{1+u}{2} 
- \frac{1-u}{2} \ln \frac{1-u}{2} 
\nonumber \\ &&
+ \frac{1+u}{2} \ln Q_+
+ \frac{1-u}{2} \ln Q_-
\label{243}
\end{eqnarray}
with
\begin{eqnarray}
\bar \xi_k &= & e^{-\mu + \mu \eta_k} f'(\xi_k)
\nonumber \\ 
\xi_k &= &
 \frac{1+u}{2} \langle \sigma_3(k) \rangle_+
+ \frac{1-u}{2} \langle \sigma_3(k) \rangle_-
\nonumber \\ 
\bar \eta_k &= & \mu e^{-\mu + \mu \eta_k} f(\xi_k)
\nonumber \\ 
\eta_k &= &
 \frac{1+u}{2} \langle \sigma_1(k) \rangle_+
+ \frac{1-u}{2} \langle \sigma_1(k) \rangle_-
\label{244}
\end{eqnarray}
In Appendix G we evaluate these expressions where the
probability distribution is large, in the Gaussian
central region.

\subsection{The Eigen model distribution function in the
Gaussian central region}

\begin{figure}[t!]
\epsfig{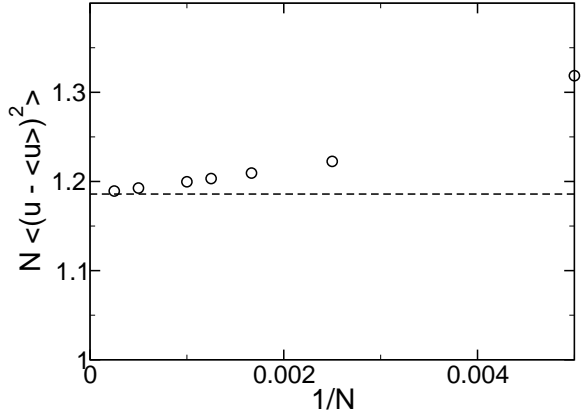}
\caption{The $O(1/N)$ shift in the variance of the
surface magnetization is
shown (circles).  Also shown is the prediction 
from Eq.\ (\ref{255}) (dashed line).  We use
$f(m) = k m^2/2 + 1$ and $d(m) = 0$ with
$k/\mu = 2$ and $\mu = 5$.
\label{fig4}
}
\end{figure}

The probability distribution function for the Eigen
model becomes Gaussian in the large $N$ limit.
Using the results from Appendix G, we find that Eq.\ (\ref{243}) becomes
\begin{eqnarray}
\frac{\ln P(u, t) }{N} &=&
({\rm const}) - \frac{1}{2} \delta u^2 \left( \frac{f'(u_*) - d'(u_*)}
{2 \mu u_* f(u_*)}
\right)
\label{253}
\end{eqnarray}
We, therefore, conclude that
\begin{eqnarray}
\left\langle \left ( u - \langle u \rangle \right)^2 \right\rangle =
 \frac{2 \mu u_* f(u_*)}{N [f'(u_*) - d'(u_*)]}
\label{254}
\end{eqnarray}
From Eqs.\ (\ref{202}), (\ref{84}), and (\ref{254}) we find
the shift of the average magnetization to be
\begin{eqnarray}
\langle u - u_* \rangle  &=&
\frac{1}{N f'(u_*)}\bigg[
b' \left [
1 - \left[ 1 - c'/{b'}    \right]^{1/2} \right]
- \frac{\mu u_* f(u_*) f''(u_*)}{f'(u_*) - d'(u_*)}
\bigg]
\label{255}
\end{eqnarray}
Shown in Figure \ref{fig4} is the comparison between
this analytical result and a numerical calculation
following the algorithm in \cite{ei89}.

\section{Conclusion}

We have derived exact functional integral representations of the
Crow-Kimura and Eigen models of quasispecies theory.
The functional integral representation of these quasispecies theories
is quite convenient because it allows us to obtain the exact
infinite genome solution of these models as well as to obtain
the finite length genome corrections.
These exact results allow us to discuss the phase transitions
that occur in these models as a function of mutation rate.
The coherent states derivation of this functional integral also
allows us to compute the full time-dependent probability distribution
function of the population in sequence space, including the dependence
on initial and final conditions.

The functional integral representation leads to an
interacting field theory for spin and mutation fields.  In
the limit of long genomes, the field theory becomes Gaussian.
We have evaluated the theory in the infinite genome limit to give the
mean replication rate for arbitrary replication and
degradation functions.  
These results were used to exhibit the phase transitions that occur
in quasispecies theory as a function of mutation rate.
For smooth replication
and degradation rate functions, we have evaluated the 
$O(1/N)$ corrections to the mean replication rate.
 We have also derived
the finite, $O(1/\sqrt{N})$ width of the virus population 
in genome space in this limit.

The functional integral representation can be
applied to generalizations of the quasispecies theories
considered here.
The extension of the present results to
arbitrary replication and degradation functions
that depend on distances from $K$ points in the space
of all possible genomes is straightforward with
the introduction of overlap parameters \cite{Deem2006}.
With the Schwinger spin coherent state formalism, the
extension to genomes with alphabets larger than binary
\cite{Burger} is also straightforward, since the
field theory in ${\bf z}^*$ and ${\bf z}$ remains Gaussian
due to constraint (\ref{3a}).

\begin{acknowledgements}
This work has been supported by DARPA \#HR00110510057.
\end{acknowledgements}

\bibliography{schwinger}

\section*{Appendix A}

To evaluate the partition function of the parallel model,
we introduce $\xi_k = \frac{1} {N} \sum_{j=1}^N {\vec z}^*_{k} (j)
\sigma_3  {\vec z}_{k-1}(j)$ with the representation
\begin{eqnarray}
&& \int {\cal D} \xi \prod_{k=1}^M \delta
\left[\xi_k - \frac{1} {N} \sum_{j=1}^N
 {\vec z}^*_{k} (j)
\sigma_3  {\vec z}_{k-1}(j) \right]
\nonumber \\ && =
 \int \left[ \prod_{k=1}^M  \frac{d \bar \xi_k d \xi_k }{2 \pi}
\right]
e^{i \sum_{k=1}^M \bar \xi_k \xi_k
- \frac{i}{N}\sum_{k=1}^M \vec z^*_k(j) \bar \xi_k \sigma_3 \vec z_{k-1}(j)}
\nonumber \\ && =
 \int \left[ \prod_{k=1}^M  \frac{i \epsilon N d \bar \xi_k d \xi_k }{2 \pi}
\right]
e^{-\epsilon N \sum_{k=1}^M \bar \xi_k \xi_k
+ \epsilon \sum_{k=1}^M \vec z^*_k(j) \bar \xi_k \sigma_3 \vec z_{k-1}(j)}
\label{22a}
\end{eqnarray}
We thus find
\begin{eqnarray}
{ Z} &=&  \lim_{M \to \infty}
 \int \left[ \prod_{k=1}^M  \frac{i \epsilon N d \bar \xi_k d \xi_k }{2 \pi}
\right]
e^{\epsilon \sum_{k=1}^M 
[ N f(\xi_k) - N \bar \xi_k \xi_k - N \mu + \Delta f] }
\nonumber \\ &&
\times
\int
 {\cal D} \lambda
 {\cal D} \vec z^*
 {\cal D} \vec z
\prod_{j=1}^N
e^{-i \lambda_j}
e^{- \sum_{k,l=1}^M \vec z_k^*(j) S(j)_{kl} \vec z_l(j) } \vert_{
\{ \vec z_0 \} = \{ e^{i \lambda} \vec z_M \} }
\nonumber \\  &=&  
\lim_{M \to \infty}
 \int \left[ \prod_{k=1}^M  \frac{i \epsilon N d \bar \xi_k d \xi_k }{2 \pi}
\right]
e^{\epsilon \sum_{k=1}^M 
[ N f(\xi_k) - N \bar \xi_k \xi_k - N \mu + \Delta f] }
\nonumber \\ &&
\times
\int
 {\cal D} \lambda
\prod_{j=1}^N 
e^{-i \lambda_j}
[\det S(j)]^{-1}
\label{23}
\end{eqnarray}
The matrix $S(j)$ has the form
\begin{eqnarray}
S(j) = 
\left( \begin{array}{ccccc} 
I & 0 & 0 &\ldots & - e^{i \lambda_j} A_1(j) \\
- A_2(j) & I & 0 & \ldots &0 \\
0 & - A_3(j) & I & \ldots &0\\
&&&\ddots\\
0&& \ldots & -A_M(j) & I
\end{array} \right) 
\label{24}
\end{eqnarray}
where $A_k(j) =  I + 
\epsilon \mu \sigma_1 + \epsilon \bar \xi_k \sigma_3$.
We find
\begin{eqnarray}
\det S(j) &=& \det \left[ I - e^{i \lambda_j} \prod_{k=1}^M A_k(j) \right]
\nonumber \\ &=&
\det\left[ I - e^{i \lambda_j} \hat T e^{\epsilon \sum_{k=1}^M
[ \mu \sigma_1 + \bar \xi_k \sigma_3 ] } \right]
\nonumber \\ &=&
e^{{\rm Tr} \ln \left[ I - {e^{i \lambda_j}}
\hat T e^{\epsilon \sum_{k=1}^M
[ \mu \sigma_1 + \bar \xi_k \sigma_3 ]} \right]}
\label{25}
\end{eqnarray}
Here the operator $\hat T$ indicates time ordering.
The partition function becomes
\begin{eqnarray}
{ Z} &=& \lim_{M \to \infty}
 \int \left[ \prod_{k=1}^M  \frac{i \epsilon N d \bar \xi_k d \xi_k }{2 \pi}
\right]
e^{\epsilon \sum_{k=1}^M 
[ N f(\xi_k) - N \bar \xi_k \xi_k - N \mu + \Delta f] }
\nonumber \\ &&
\times
\int
 {\cal D} \lambda
\prod_{j=1}^N e^{-i \lambda_j}
e^{- {\rm Tr} \ln \left[ I - {e^{i \lambda_j}}
\hat T e^{\epsilon \sum_{k=1}^M
[ \mu \sigma_1 + \bar \xi_k \sigma_3 ]} \right]}
\nonumber  \\
&=& \lim_{M \to \infty}
 \int \left[ \prod_{k=1}^M  \frac{i \epsilon N d \bar \xi_k d \xi_k }{2 \pi}
\right]
e^{\epsilon \sum_{k=1}^M 
[ N f(\xi_k) - N \bar \xi_k \xi_k - N \mu + \Delta f] }
\nonumber \\ &&
\times
\prod_{j=1}^N Q(j)
\label{26}
\end{eqnarray}
where
\begin{eqnarray}
Q(j) &=&
 {\rm Tr}
\prod_{k=1}^M [ I + \epsilon \mu \sigma_1 + \epsilon  \bar \xi_k \sigma_3]
\nonumber \\ &\sim&
 {\rm Tr}
 \hat T e^{\epsilon \sum_{k=1}^M
[ \mu \sigma_1 + \bar \xi_k \sigma_3 ] }
\nonumber \\ &=&
{\rm Tr} \hat T e^{ \int_0^t  dt'
[ \mu \sigma_1 + \bar \xi(t') \sigma_3 ] }
\label{27}
\end{eqnarray}

\section*{Appendix B}

In this Appendix we evaluate the fluctuation corrections to the
parallel model.
The second term in the last derivative of Eq.\ (\ref{34}) is given by
\begin{eqnarray}
 \frac{
 {\rm Tr}  \sigma_3
e^{\epsilon M (\mu \sigma_1 + \bar \xi_c \sigma_3)} 
}
{
 {\rm Tr} e^{\epsilon M(\mu \sigma_1 + \bar \xi_c \sigma_3)} 
}
&= &
\frac{\bar \xi_c}{ [\mu^2 + \bar \xi_c^2]^{1/2}}
 \tanh t [\mu^2 + \bar \xi_c^2]^{1/2}
\nonumber \\ &\sim&
\frac{\bar \xi_c}{ [\mu^2 + \bar \xi_c^2]^{1/2}}
{\rm ~as~} t \to \infty
\label{35}
\end{eqnarray}
The first term in the last derivative is given for $k \ne l$ by
\begin{eqnarray}
&&
 \frac{
 {\rm Tr}
  \sigma_3
e^{\epsilon \vert l-k \vert (\mu \sigma_1 + \bar \xi_c \sigma_3)} 
  \sigma_3
e^{\epsilon (M-\vert l-k \vert )(\mu \sigma_1 + \bar \xi_c \sigma_3)} 
}
{
 {\rm Tr} e^{\epsilon M(\mu \sigma_1 + \bar \xi_c \sigma_3)} 
}
\nonumber \\ && =
\frac{\bar \xi_c^2}{ \mu^2 + \bar \xi_c^2}
 + 
\frac{\mu^2}{ \mu^2 + \bar \xi_c^2}
\frac{
\cosh (t_1 - t_2) \sqrt{\mu^2 + \bar \xi_c^2}
} 
{
\cosh t \sqrt{\mu^2 + \bar \xi_c^2}
}
\nonumber \\ && \sim
\frac{\bar \xi_c^2}{ \mu^2 + \bar \xi_c^2}
 + 
\frac{\mu^2}{ \mu^2 + \bar \xi_c^2}
e^{(\vert t_1 - t_2 \vert - t) \sqrt{\mu^2 + \bar \xi_c^2} }
{\rm ~as~} t \to \infty
\label{36}
\end{eqnarray}
where 
$t_1 = \epsilon \vert l-k \vert$ and $t_2 = \epsilon (M-\vert l-k \vert )$.
For $k=l$, 
the first term in the last derivative is zero from the
first line of Eq.\ (\ref{27}).
We find
\begin{eqnarray}
\frac{\partial^2 S}{\partial \bar \xi_k \partial \bar \xi_{l}} 
= -N \epsilon^2 
\frac{\mu^2}{ \mu^2 + \bar \xi_c^2}
M_{kl} + N \epsilon^2 \delta_{kl}
\label{37}
\end{eqnarray}
where 
\begin{eqnarray}
M_{k l} = e^{\epsilon [\vert M - 2 \vert l-k \vert \vert - M] 
\sqrt{\mu^2 + \bar \xi_c^2}}
\label{38}
\end{eqnarray}
We let $\bar \eta_k = \delta \bar \xi_k  \sqrt(\epsilon N)$
and
$ \eta_k = \delta \xi_k  \sqrt(\epsilon N)$, and the
partition function becomes
\begin{eqnarray}
{ Z} &=& 
e^{-S(\bar \xi_c, \xi_c)}
\lim_{M \to \infty}
 \int \left[ \prod_{k=1}^M  \frac{i d \bar \eta_k d \eta_k }{2 \pi}
\right]
e^{\frac{1}{2} \sum_k [f''(\xi_c) \eta_k^2 - 2 \bar \eta_k \eta_k]
}
\nonumber \\ && \times e^{ \frac{1}{2} \sum_{k,l} 
\left[ \epsilon 
\frac{\mu^2}{ \mu^2 + \bar \xi_c^2} 
M_{k l}
\bar \eta_k \bar \eta_{l}
- \epsilon 
\bar \eta_k \bar \eta_{l} \delta_{kl} \right] }
\label{39}
\end{eqnarray}
We let $\eta_k' = i \eta_k \sqrt{f''(\xi_c)}$
and $\bar \eta_k' = \bar \eta_k/ \sqrt{f''(\xi_c)}$ to 
obtain
\begin{eqnarray}
{ Z} &=& 
e^{-S(\bar \xi_c, \xi_c)}
\lim_{M \to \infty}
 \int \left[ \prod_{k=1}^M  \frac{d \bar \eta_l' d \eta_l' }{2 \pi}
\right]
e^{-\frac{1}{2} \sum_k {\eta_l'}^2 
+ i \sum_k\bar \eta_l' \eta_l'
}
\nonumber \\ && \times e^{ \frac{1}{2} \sum_{k,l} 
\left[\epsilon 
\frac{\mu^2 f''(\xi_c)}{ \mu^2 + \bar \xi_c^2} 
M_{k l}
\bar \eta_k' \bar \eta_{l}'
- \epsilon 
f''(\xi_c) \bar \eta_k' \bar \eta_{l}' \delta_{kl} \right] }
\label{40}
\end{eqnarray}
Integrating over $\eta'$, we find
\begin{eqnarray}
{ Z} &=& 
e^{-S(\bar \xi_c, \xi_c)}
\lim_{M \to \infty}
 \int \left[ \prod_{k=1}^M  \frac{d \bar \eta_l' }{\sqrt{2 \pi}}
\right]
e^{ -\frac{1}{2} \sum_{k,l} \left[
\delta_{k l}
- \epsilon 
\frac{\mu^2 f''(\xi_c)}{ \mu^2 + \bar \xi_c^2} 
M_{k l}
+ \epsilon f''(\xi_c) \delta_{kl}
\right]
\bar \eta_k' \bar \eta_{l}'
}
\nonumber \\ &=&
e^{-S(\bar \xi_c, \xi_c)}
\lim_{M \to \infty}
 \int \left[ \prod_{k=1}^M  \frac{d \bar \eta_l' }{\sqrt{2 \pi}}
\right]
e^{ -\frac{1}{2} \sum_{k,l}
\bar \eta_l' F_{k l} \bar \eta_{l}'
}
\nonumber \\ &=&
e^{-S(\bar \xi_c, \xi_c)}
( \det F ) ^{-1/2}
\label{41}
\end{eqnarray}
where 
\begin{eqnarray}
F_{k l} &= & \delta_{k l} - \epsilon
\frac{\mu^2 f''(\xi_c)}{ \mu^2 + \bar \xi_c^2}
M_{k l} + \epsilon f''(\xi_c) \delta_{kl}
\nonumber \\ &=&
 \delta_{k l} - \epsilon \frac{a}{b^2} M_{kl}
 + \epsilon f''(\xi_c) \delta_{kl}
\label{42}
\end{eqnarray}
where $a = \mu^2 f''(\xi_c)$ and $b^2 = \mu^2 + \bar \xi_c^2$.
We define
$ F'$, where $F_{kl} = F_{kl}' + \epsilon f''(\xi_c) \delta_{kl} $.
We note that 
\begin{eqnarray}
\det F &=& 
\det[I + \epsilon f''(\xi_c) I - \epsilon \frac{a }{b^2} M]
\nonumber \\ &=&
(\det [I + \epsilon f''(\xi_c) I])
(\det [I - \epsilon \frac{a }{b^2} M [I + \epsilon f''(\xi_c) I]^{-1}) 
\nonumber \\ &\sim&
(\det [I + \epsilon f''(\xi_c) I])
(\det [I - \epsilon \frac{a }{b^2} M])
\nonumber \\ &=&
(\det F') 
e ^{ {\rm Tr}~ \ln [ I + \epsilon f''(\xi_c) ] }
\nonumber \\ &\sim&
(\det F') 
e ^{ {\rm Tr}~ \epsilon f''(\xi_c) I  }
\nonumber \\ &=&
(\det F') 
e ^{ t f''(\xi_c)   }
\label{42d}
\end{eqnarray}
To evaluate $\det F'$ we use Fourier space.  
We find
\begin{eqnarray}
\hat F'(k) = 
1 - \frac{4 \epsilon^2 a}{ b}
\frac{1}{4 b^2 \epsilon^2 + k^2}
\label{42a}
\end{eqnarray}
and
\begin{eqnarray}
\det F' &=& 
\prod_i \lambda_i = \prod_k \hat F'(k)
\label{42b}
\end{eqnarray}
where $k = 2 \pi n / M = 2 \pi \epsilon n / t$, $n = - M/2, \ldots,
(M-1)/2$.  In the 
limit of infinite $M$, we find
\begin{eqnarray}
\det F' &=& 
\frac{
\left( 
\sinh t \sqrt{b^2 - a/b}
\right)^2
}{
\left( 
\sinh b t
\right)^2
}
\nonumber \\ &=&
\frac{ \left(
\sinh t \left[ \mu^2 + \bar \xi_c^2 - 
    \mu^2 f''(\xi_c) / \sqrt{\mu^2 + \bar \xi_c^2} \right]^{1/2}
 \right)^2 }
{ \left( \sinh t \sqrt{\mu^2 + \bar \xi_c^2} \right)^2 } 
\nonumber \\ 
&\sim &
e^{-2 t \left[ \sqrt{\mu^2 + \bar \xi_c^2} - 
[\mu^2 + \bar \xi_c^2 - 
    \mu^2 f''(\xi_c) / \sqrt{\mu^2 + \bar \xi_c^2}]^{1/2} \right]^{1/2} }
{\rm ~as~} t \to \infty
\label{42c}
\end{eqnarray}
Using Eq.\ (\ref{32}), (\ref{41}), (\ref{42c}), and (\ref{42d})
 we find the mean replication rate at large time becomes
\begin{eqnarray}
\frac{\ln Z}{t N} &=&
f(\xi_c) - \bar \xi_c \xi_c + \sqrt{ \mu^2 + \bar \xi_c^2}
- \mu +
\frac{\Delta f}{N} 
- \frac{ f''(\xi_c)}{2N} 
\nonumber \\ && +
\frac{1}{N} \frac{\mu}{ \sqrt{ 1 - \xi_c^2}}
\left[1 - \left[1 -  f''(\xi_c) (1-\xi_c^2)^{3/2} / \mu\right]^{1/2} \right]
\label{44}
\end{eqnarray}
with
\begin{eqnarray}
\bar \xi_c &=& f'(\xi_c)
\nonumber \\ 
\xi_c &=& \frac{\bar \xi_c}{ [\mu^2 + \bar \xi_c^2]^{1/2}}
 \tanh t [\mu^2 + \bar \xi_c^2]^{1/2}
\label{45}
\end{eqnarray}

\section*{Appendix C}

To evaluate the partition function of the Eigen model,
we define magnetization and mutation fields with 
$\xi_k = \frac{1} {N} \sum_{j=1}^N {\vec z}^*_{k} (j)
\sigma_3  {\vec z}_{k-1}(j)$ 
and
$\eta_k = \frac{1} {N} \sum_{j=1}^N {\vec z}^*_{k} (j)
\sigma_1  {\vec z}_{k-1}(j)$ 
to write the partition function for the Eigen model as
\begin{eqnarray}
{ Z} &=&  \lim_{M \to \infty}
 \int \left[ \prod_{k=1}^M  \frac{i \epsilon N d \bar \xi_k d \xi_k }{2 \pi}
\right]
 \left[ \prod_{k=1}^M  \frac{i \epsilon N d \bar \eta_k d \eta_k }{2 \pi}
\right]
\nonumber \\ &&
\times e^{\epsilon N \sum_{k=1}^M 
[  e^{-\mu_1} e^{ \mu \eta_k}
f(\xi_k)  - d(\xi_k) -  \bar \xi_k \xi_k
- \bar \eta_k \eta_k
] }
\nonumber \\ &&
\times
\int
 {\cal D} \lambda
 {\cal D} \vec z^*
 {\cal D} \vec z
\prod_{j=1}^N
e^{-i \lambda_j}
e^{- \sum_{k,l=1}^M \vec z_k^*(j) S(j)_{kl} \vec z_l(j) } \vert_{
\{ \vec z_0 \} = \{ e^{i \lambda} \vec z_M \} }
\nonumber \\  &=&  
\lim_{M \to \infty}
 \int \left[ \prod_{k=1}^M  \frac{i \epsilon N d \bar \xi_k d \xi_k }{2 \pi}
\right]
 \left[ \prod_{k=1}^M  \frac{i \epsilon N d \bar \eta_k d \eta_k }{2 \pi}
\right]
\nonumber \\ &&
e^{\epsilon N \sum_{k=1}^M 
[  e^{-\mu_1} e^{ \mu \eta_k}
f(\xi_k)  - d(\xi_k) -  \bar \xi_k \xi_k
- \bar \eta_k \eta_k
] }
\nonumber \\ &&
\times
\int
[ {\cal D} \lambda]
e^{-i \lambda_j}
\prod_{j=1}^N [\det S(j)]^{-1}
\label{51}
\end{eqnarray}
The matrix $S(j)$ has the form of Eq.\ (\ref{24})
with $A_k(j) =  I + 
\epsilon \bar \eta_k \sigma_1 + \epsilon \bar \xi_k \sigma_3$.
The determinant and constraint evaluate as
in Eqs. (\ref{25}--\ref{26}) to give
\begin{eqnarray}
{ Z} = \lim_{M \to \infty}
 \int&&
 \left[ \prod_{k=1}^M  \frac{i \epsilon N d \bar \xi_k d \xi_k }{2 \pi}
\right]
 \left[ \prod_{k=1}^M  \frac{i \epsilon N d \bar \eta_k d \eta_k }{2 \pi}
\right]
\nonumber \\ &&
\times e^{\epsilon N \sum_{k=1}^M 
[  e^{-\mu_1} e^{ \mu \eta_k}
f(\xi_k)  - d(\xi_k) -  \bar \xi_k \xi_k
- \bar \eta_k \eta_k
] }
\nonumber \\ &&
\times
\prod_{j=1}^N Q(j)
\label{52}
\end{eqnarray}
where
\begin{eqnarray}
Q(j) &=& 
 {\rm Tr} 
\prod_{k=1}^N
[ I + \epsilon \bar \eta_k \sigma_1 + \epsilon \bar \xi_k \sigma_3 ] 
\nonumber \\ &\sim&
 {\rm Tr} 
\hat T e^{\epsilon \sum_{k=1}^M
[\bar \eta_k \sigma_1 + \bar \xi_k \sigma_3 ] }
\nonumber \\ &=&
{\rm Tr} \hat T e^{ \int_0^t  dt'
[ \bar \eta(t') \sigma_1 + \bar \xi(t') \sigma_3 ] }
\label{53}
\end{eqnarray}

\section*{Appendix D}
We here determine the fluctuation corrections 
to the mean excess replication rate per site,
$(\ln Z)/(t N)$, in the Eigen model.
We expand the action around the saddle point limit
\begin{eqnarray}
S[  \bar \xi,  \xi,
 \bar \eta,  \eta ] &=& 
S(\bar \xi_c, \xi_c,
 \bar \eta_c, \eta_c )  +
\frac{1}{2} \sum_{k,l=1}^M  \bigg[
\left. \frac{\partial^2 S}{\partial \xi_k \partial \xi_{l}} 
 \right\vert_{\bar \xi_c, \xi_c, \bar \eta_c, \eta_c}
\delta \xi_k \delta \xi_{l}
\nonumber \\ &&
+ 2 \left. \frac{\partial^2 S}{\partial \xi_k \partial \bar \xi_{l}} 
 \right\vert_{\bar \xi_c, \xi_c, \bar \eta_c, \eta_c}
\delta \xi_k \delta \bar \xi_{l}
+ \left. \frac{\partial^2 S}{\partial \bar \xi_k \partial \bar \xi_{l}} 
 \right\vert_{\bar \xi_c, \xi_c, \bar \eta_c, \eta_c}
\delta \bar \xi_k \delta \bar \xi_{l}
\nonumber \\ &&
+ \left. \frac{\partial^2 S}{\partial \eta_k \partial \eta_{l}} 
 \right\vert_{\bar \xi_c, \xi_c, \bar \eta_c, \eta_c}
\delta \eta_k \delta \eta_{l}
+ 2 \left. \frac{\partial^2 S}{\partial \eta_k \partial \bar \eta_{l}} 
 \right\vert_{\bar \xi_c, \xi_c, \bar \eta_c, \eta_c}
\delta \eta_k \delta \bar \eta_{l}
\nonumber \\ &&
+ \left. \frac{\partial^2 S}{\partial \bar \eta_k \partial \bar \eta_{l}} 
 \right\vert_{\bar \xi_c, \xi_c, \bar \eta_c, \eta_c}
\delta \bar \eta_k \delta \bar \eta_{l}
\nonumber \\ &&
+ 2 \left. \frac{\partial^2 S}{\partial \xi_k \partial \bar \eta_{l}} 
 \right\vert_{\bar \xi_c, \xi_c, \bar \eta_c, \eta_c}
\delta \xi_k \delta \bar \eta_{l}
+ 2 \left. \frac{\partial^2 S}{\partial \eta_k \partial \bar \xi_{l}} 
 \right\vert_{\bar \xi_c, \xi_c, \bar \eta_c, \eta_c}
\delta \eta_k \delta \bar \xi_{l}
\nonumber \\ &&
+ 2 \left. \frac{\partial^2 S}{\partial \xi_k \partial  \eta_{l}} 
 \right\vert_{\bar \xi_c, \xi_c, \bar \eta_c, \eta_c}
\delta \xi_k \delta  \eta_{l}
+ 2 \left. \frac{\partial^2 S}{\partial \bar \eta_k \partial \bar \xi_{l}} 
 \right\vert_{\bar \xi_c, \xi_c, \bar \eta_c, \eta_c}
\delta \bar \eta_k \delta \bar \xi_{l}
\bigg]
\nonumber \\
\label{70}
\end{eqnarray}
We find
\begin{eqnarray}
\left. \frac{\partial^2 S}{\partial \xi_k \partial \xi_{l}}  
 \right\vert_{\bar \xi_c, \xi_c, \bar \eta_c, \eta_c}
&=&
-N \epsilon e^{-mu_1 + \mu \eta_c} f''(\xi_c) \delta_{k l}
+ N \epsilon d''(\xi_c) \delta_{kl}
\nonumber \\
\left. \frac{\partial^2 S}{\partial \xi_k \partial \eta_{l}}  
 \right\vert_{\bar \xi_c, \xi_c, \bar \eta_c, \eta_c}
&=&
-N \epsilon \mu e^{-mu_1 + \mu \eta_c} f'(\xi_c) \delta_{k l}
\nonumber \\
\left. \frac{\partial^2 S}{\partial \eta_k \partial \eta_{l}}  
 \right\vert_{\bar \xi_c, \xi_c, \bar \eta_c, \eta_c}
&=&
-N \epsilon \mu^2 e^{-mu_1 + \mu \eta_c} f(\xi_c) \delta_{k l}
\nonumber \\
\left. \frac{\partial^2 S}{\partial \xi_k \partial \bar \xi_{l}} 
 \right\vert_{\bar \xi_c, \xi_c, \bar \eta_c, \eta_c}
 &=&
 \epsilon N \delta_{k l}
\nonumber \\
\left. \frac{\partial^2 S}{\partial \eta_k \partial \bar \eta_{l}}  
 \right\vert_{\bar \xi_c, \xi_c, \bar \eta_c, \eta_c}
&=&
 \epsilon N \delta_{k l}
\nonumber \\
\left. \frac{\partial^2 S}{\partial \bar \xi_k \partial \bar \xi_{l}} 
 \right\vert_{\bar \xi_c, \xi_c, \bar \eta_c, \eta_c}
 &=&
- N \epsilon^2
 \frac{
 {\rm Tr}
  \sigma_3
e^{\epsilon \vert l-k \vert (\bar \eta_c \sigma_1 + \bar \xi_c \sigma_3)} 
  \sigma_3
e^{\epsilon (M-\vert l-k \vert (\bar \eta_c \sigma_1 + \bar \xi_c \sigma_3)} 
}
{
 {\rm Tr} e^{\epsilon M( \bar \eta_c \sigma_1 + \bar \xi_c \sigma_3)} 
} (1-\delta_{kl})
\nonumber \\ && +
N \epsilon^2 \left(
 \frac{
 {\rm Tr}  \sigma_3
e^{\epsilon M (\bar \eta_c \sigma_1 + \bar \xi_c \sigma_3)} 
}
{
 {\rm Tr} e^{\epsilon M(\bar \eta_c \sigma_1 + \bar \xi_c \sigma_3)} 
}
\right)^2
\nonumber \\
\left. \frac{\partial^2 S}{\partial \bar \eta_k \partial \bar \eta_{l}}  
 \right\vert_{\bar \xi_c, \xi_c, \bar \eta_c, \eta_c}
&=&
- N \epsilon^2
 \frac{
 {\rm Tr}
  \sigma_1
e^{\epsilon \vert l-k \vert (\bar \eta_c \sigma_1 + \bar \eta_c \sigma_3)} 
  \sigma_1
e^{\epsilon (M-\vert l-k \vert (\bar \eta_c \sigma_1 + \bar \eta_c \sigma_3)} 
}
{
 {\rm Tr} e^{\epsilon M(\bar \eta_c \sigma_1 + \bar \eta_c \sigma_3)} 
} (1-\delta_{kl})
\nonumber \\ && +
N \epsilon^2 \left(
 \frac{
 {\rm Tr}  \sigma_1
e^{\epsilon M (\bar \eta_c \sigma_1 + \bar \eta_c \sigma_3)} 
}
{
 {\rm Tr} e^{\epsilon M(\bar \eta_c \sigma_1 + \bar \eta_c \sigma_3)} 
}
\right)^2
\nonumber \\
\left.
\frac{\partial^2 S}{\partial \bar \eta_k \partial \bar \xi_{l}}
 \right\vert_{\bar \xi_c, \xi_c, \bar \eta_c, \eta_c, l \ge k}
 &=&
- N \epsilon^2
 \frac{
 {\rm Tr}
  \sigma_3
e^{\epsilon ( l-k ) (\bar \eta_c \sigma_1 + \bar \xi_c \sigma_3)} 
  \sigma_1
e^{\epsilon (M- l+k ) (\bar \eta_c \sigma_1 + \bar \xi_c \sigma_3)} 
}
{
 {\rm Tr} e^{\epsilon M(\bar \eta_c \sigma_1 + \bar \xi_c \sigma_3)} 
} (1-\delta_{kl})
\nonumber \\ && +
N \epsilon^2 
\left(
 \frac{
 {\rm Tr}  \sigma_1
e^{\epsilon M (\bar \eta_c \sigma_1 + \bar \xi_c \sigma_3)} 
}
{
 {\rm Tr} e^{\epsilon M(\bar \eta_c \sigma_1 + \bar \xi_c \sigma_3)} 
}
\right)
\left(
 \frac{
 {\rm Tr}  \sigma_3
e^{\epsilon M (\bar \eta_c \sigma_1 + \bar \xi_c \sigma_3)} 
}
{
 {\rm Tr} e^{\epsilon M(\bar \eta_c \sigma_1 + \bar \xi_c \sigma_3)} 
}
\right)
\nonumber \\ 
\left.
\frac{\partial^2 S}{\partial \bar \eta_k \partial \bar \xi_{l}}
 \right\vert_{\bar \xi_c, \xi_c, \bar \eta_c, \eta_c, l \le k}
&=&
- N \epsilon^2
 \frac{
 {\rm Tr}
  \sigma_1
e^{\epsilon ( k-l) (\bar \eta_c \sigma_1 + \bar \xi_c \sigma_3)} 
  \sigma_3
e^{\epsilon (M- k+l ) (\bar \eta_c \sigma_1 + \bar \xi_c \sigma_3)} 
}
{
 {\rm Tr} e^{\epsilon M(\bar \eta_c \sigma_1 + \bar \xi_c \sigma_3)} 
} (1-\delta_{kl})
\nonumber \\ && +
N \epsilon^2 
\left(
 \frac{
 {\rm Tr}  \sigma_1
e^{\epsilon M (\bar \eta_c \sigma_1 + \bar \xi_c \sigma_3)} 
}
{
 {\rm Tr} e^{\epsilon M(\bar \eta_c \sigma_1 + \bar \xi_c \sigma_3)} 
}
\right)
\left(
 \frac{
 {\rm Tr}  \sigma_3
e^{\epsilon M (\bar \eta_c \sigma_1 + \bar \xi_c \sigma_3)} 
}
{
 {\rm Tr} e^{\epsilon M(\bar \eta_c \sigma_1 + \bar \xi_c \sigma_3)} 
}
\right)
\nonumber \\
\label{71}
\end{eqnarray}
The other two derivatives in Eq.\ (\ref{70}) are zero.
We find
\begin{eqnarray}
 \frac{
 {\rm Tr}  \sigma_3
e^{\epsilon M (\bar \eta_c \sigma_1 + \bar \xi_c \sigma_3)} 
}
{
 {\rm Tr} e^{\epsilon M(\bar \eta_c \sigma_1 + \bar \xi_c \sigma_3)} 
}
&= &
\frac{\bar \xi_c}{ [\bar \eta_c^2 + \bar \xi_c^2]^{1/2}}
 \tanh t [\bar \eta_c^2 + \bar \xi_c^2]^{1/2}
\nonumber \\ &\sim&
\frac{\bar \xi_c}{ [\bar \eta_c^2 + \bar \xi_c^2]^{1/2}}
{\rm ~as~} t \to \infty
\label{71a}
\end{eqnarray}
and
\begin{eqnarray}
&&
 \frac{
 {\rm Tr}
  \sigma_3
e^{\epsilon \vert l-k \vert (\bar \eta_c \sigma_1 + \bar \xi_c \sigma_3)} 
  \sigma_3
e^{\epsilon (M-\vert l-k \vert )(\bar \eta_c \sigma_1 + \bar \xi_c \sigma_3)} 
}
{
 {\rm Tr} e^{\epsilon M(\bar \eta_c \sigma_1 + \bar \xi_c \sigma_3)} 
}
\nonumber \\ && =
\frac{\bar \xi_c^2}{ \bar \eta_c^2 + \bar \xi_c^2}
 + 
\frac{\bar \eta_c^2}{ \bar \eta_c^2 + \bar \xi_c^2}
\frac{
\cosh (t_1 - t_2) \sqrt{\bar \eta_c^2 + \bar \xi_c^2}
} 
{
\cosh t \sqrt{\bar \eta_c^2 + \bar \xi_c^2}
}
\nonumber \\ && \sim
\frac{\bar \xi_c^2}{ \bar \eta_c^2 + \bar \xi_c^2}
 + 
\frac{\bar \eta_c^2}{ \bar \eta_c^2 + \bar \xi_c^2}
e^{(\vert t_1 - t_2 \vert - t) \sqrt{\bar \eta_c^2 + \bar \xi_c^2} }
{\rm ~as~} t \to \infty
\label{71b}
\end{eqnarray}
where 
$t_1 = \epsilon \vert l-k \vert$ and $t_2 = \epsilon (M-\vert l-k \vert )$.
We find
\begin{eqnarray}
\frac{\partial^2 S}{\partial \bar \xi_k \partial \bar \xi_{l}} 
= -N \epsilon^2 
\frac{\bar \eta_c^2}{ \bar \eta_c^2 + \bar \xi_c^2}
M'_{kl} + N \epsilon^2 \delta_{kl}
\label{71c}
\end{eqnarray}
where 
\begin{eqnarray}
M'_{k l} = e^{\epsilon [\vert M - 2 \vert l-k \vert \vert - M] 
\sqrt{\bar \eta_c^2 + \bar \xi_c^2}}
\label{71d}
\end{eqnarray}
The other traces evaluate as
\begin{eqnarray}
 \frac{
 {\rm Tr}  \sigma_1
e^{\epsilon M (\bar \eta_c \sigma_1 + \bar \xi_c \sigma_3)} 
}
{
 {\rm Tr} e^{\epsilon M(\bar \eta_c \sigma_1 + \bar \xi_c \sigma_3)} 
}
&= &
\frac{\bar \eta_c}{ [\bar \eta_c^2 + \bar \xi_c^2]^{1/2}}
 \tanh t [\bar \eta_c^2 + \bar \xi_c^2]^{1/2}
\nonumber \\ &\sim&
\frac{\bar \eta_c}{ [\bar \eta_c^2 + \bar \xi_c^2]^{1/2}}
{\rm ~as~} t \to \infty
\label{73}
\end{eqnarray}
and
\begin{eqnarray}
&&
 \frac{
 {\rm Tr}
  \sigma_1
e^{\epsilon \vert l-k \vert (\bar \eta_c \sigma_1 + \bar \xi_c \sigma_3)} 
  \sigma_1
e^{\epsilon (M-\vert l-k \vert )(\bar \eta_c \sigma_1 + \bar \xi_c \sigma_3)} 
}
{
 {\rm Tr} e^{\epsilon M(\bar \eta_c \sigma_1 + \bar \xi_c \sigma_3)} 
}
\nonumber \\ && =
\frac{\bar \eta_c^2}{ \bar \eta_c^2 + \bar \xi_c^2}
 + 
\frac{\xi_c^2}{ \bar \eta_c^2 + \bar \xi_c^2}
\frac{
\cosh (t_1 - t_2) \sqrt{\bar \eta_c^2 + \bar \xi_c^2}
} 
{
\cosh t \sqrt{\bar \eta_c^2 + \bar \xi_c^2}
}
\nonumber \\ && \sim
\frac{\bar \eta_c^2}{ \bar \eta_c^2 + \bar \xi_c^2}
 + 
\frac{\xi_c^2}{ \bar \eta_c^2 + \bar \xi_c^2}
e^{(\vert t_1 - t_2 \vert - t) \sqrt{\bar \eta_c^2 + \bar \xi_c^2} }
{\rm ~as~} t \to \infty
\label{74}
\end{eqnarray}
We also find
\begin{eqnarray}
&&
 \frac{
 {\rm Tr}
  \sigma_1
e^{t_1 (\bar \eta_c \sigma_1 + \bar \xi_c \sigma_3)} 
  \sigma_3
e^{t_2 (\bar \eta_c \sigma_1 + \bar \xi_c \sigma_3)} 
}
{
 {\rm Tr} e^{t (\bar \eta_c \sigma_1 + \bar \xi_c \sigma_3)} 
}
=
 \frac{
 {\rm Tr}
  \sigma_3
e^{t_2 (\bar \eta_c \sigma_1 + \bar \xi_c \sigma_3)} 
  \sigma_1
e^{t_1 (\bar \eta_c \sigma_1 + \bar \xi_c \sigma_3)} 
}
{
 {\rm Tr} e^{t (\bar \eta_c \sigma_1 + \bar \xi_c \sigma_3)} 
}
\nonumber \\ && =
\frac{\bar \eta_c \bar \xi_c}{ \bar \eta_c^2 + \bar \xi_c^2}
 - 
\frac{\bar \eta_c \bar \xi_c}{ \bar \eta_c^2 + \bar \xi_c^2}
\frac{
\cosh (t_1 - t_2) \sqrt{\bar \eta_c^2 + \bar \xi_c^2}
} 
{
\cosh t \sqrt{\bar \eta_c^2 + \bar \xi_c^2}
}
\nonumber \\ && \sim
\frac{\bar \eta_c \bar \xi_c}{ \bar \eta_c^2 + \bar \xi_c^2}
 - 
\frac{\bar \eta_c \bar \xi_c}{ \bar \eta_c^2 + \bar \xi_c^2}
e^{(\vert t_1 - t_2 \vert - t) \sqrt{\bar \eta_c^2 + \bar \xi_c^2} }
{\rm ~as~} t \to \infty
\label{75}
\end{eqnarray}
We, thus, have
\begin{eqnarray}
\frac{\partial^2 S}{\partial \bar \eta_k \partial \bar \eta_{l}} 
= -N \epsilon^2 
\frac{\bar \xi_c^2}{ \bar \eta_c^2 + \bar \xi_c^2}
M'_{kl} + N \epsilon^2 \delta_{kl}
\label{76}
\end{eqnarray}
and
\begin{eqnarray}
\frac{\partial^2 S}{\partial \bar \eta_k \partial \bar \xi_{l}} =
\frac{\partial^2 S}{\partial \bar \eta_l \partial \bar \xi_{k}} 
= N \epsilon^2 
\frac{\bar \xi_c \bar \eta_c}{ \bar \eta_c^2 + \bar \xi_c^2}
M'_{kl}
\label{77}
\end{eqnarray}
With these results, letting primes denote the
fluctuation variables, and setting $\mu_1 = \mu$,
 the partition function becomes
\begin{eqnarray}
{  Z}  &= &
e^{-S(\bar \xi_c, \xi_c,
 \bar \eta_c, \eta_c )}
\lim_{M \to \infty}
 \int
 \left[ \prod_{k=1}^M  \frac{i \epsilon N d \bar \xi_k' d \xi_k' }{2 \pi}
\right]
 \left[ \prod_{k=1}^M  \frac{i \epsilon N d \bar \eta_k' d \eta_k' }{2 \pi}
\right]
\nonumber \\ &&
\times
e^{\frac{1}{2} \epsilon N \sum_{k=1}^M 
[ e^{\mu(\eta_c -1)} f''(\xi_c) {\xi_k'}^2
- d ''(\xi_c) {\xi_k'}^2
+ 2 \mu e^{\mu(\eta_c -1)} f'(\xi_c) \xi_k' \eta_k'} ]
\nonumber \\ &&
\times
e^{\frac{1}{2} \epsilon N \sum_{k=1}^M 
 [ \mu^2 e^{\mu(\eta_c -1)}f(\xi_c) {\eta_k'}^2
- 2 \bar \xi_k' \xi_k'
- 2 \bar \eta_k' \eta_k'
] }
\nonumber \\ &&
\times 
e^{\frac{1}{2} \epsilon^2 N \sum_{k,l=1}^M 
[
\bar \eta_c^2 \bar \xi_k' \bar \xi_l'
- 2 \bar \eta_c \bar \xi_c  \bar \eta_k' \bar \xi_l'
+ \bar \xi_c^2 \bar \eta_k' \bar \eta_l'
]
\frac{M'_{kl}}{\bar \eta_c^2 + \bar \xi_c^2}
}
\nonumber \\ &&
\times 
e^{-\frac{1}{2} \epsilon^2 N \sum_{k,l=1}^M 
[
 \bar \xi_k' \bar \xi_l' + 
 \bar \eta_k' \bar \eta_l'
]
\delta_{kl} }
\label{78}
\end{eqnarray}
We remove the primes, letting 
$\bar \eta_k' = \bar \eta_k/ \sqrt{\epsilon N}$,
$ \eta_k' =  -i \eta_k/ \sqrt{\epsilon N}$,
$\bar \xi_k' = \bar \xi_k/ \sqrt{\epsilon N}$,
and
$\xi_k' = -i \xi_k/ \sqrt{\epsilon N}$,
to find
\begin{eqnarray}
{  Z}  &= &
e^{-S(\bar \xi_c, \xi_c,
 \bar \eta_c, \eta_c )}
\lim_{M \to \infty}
 \int
 \left[ \prod_{k=1}^M  \frac{ d \bar \xi_k d \xi_k }{2 \pi}
\right]
 \left[ \prod_{k=1}^M  \frac{ d \bar \eta_k d \eta_k }{2 \pi}
\right]
\nonumber \\ &&
\times
e^{-\frac{1}{2} \sum_{k=1}^M 
[ e^{\mu(\eta_c -1)} f''(\xi_c) {\xi_k}^2
- d''(\xi_c) {\xi_k}^2
+ 2 \mu e^{\mu(\eta_c -1)} f'(\xi_c) \xi_k \eta_k} ]
\nonumber \\ &&
\times
e^{-\frac{1}{2} \sum_{k=1}^M 
 [ \mu^2 e^{\mu(\eta_c -1)} f(\xi_c) {\eta_k}^2
- 2 i \bar \xi_k \xi_k
- 2 i \bar \eta_k \eta_k
] }
\nonumber \\ &&
\times 
e^{\frac{1}{2} \epsilon \sum_{k,l=1}^M 
[
\bar \eta_c^2 \bar \xi_k \bar \xi_l
- 2 \bar \eta_c \bar \xi_c  \bar \eta_k \bar \xi_l
+ \bar \xi_c^2 \bar \eta_k \bar \eta_l
]
\frac{M'_{kl}}{\bar \eta_c^2 + \bar \xi_c^2}
}
\nonumber \\ &&
\times 
e^{-\frac{1}{2} \epsilon N \sum_{k,l=1}^M 
[
 \bar \xi_k \bar \xi_l + 
 \bar \eta_k \bar \eta_l
]
\delta_{kl} }
\label{79}
\end{eqnarray}
We let $\vec x_k = (\xi_k, \eta_k)$.
We denote the summand that does not involve bars as
$-\frac{1}{2} \vec x_k^\top B^\top B \vec x_k$; that is
$B^\top B  = e^{\mu(\eta_c -1)}
 \left( \begin{array}{cc} f''(\xi_c) & \mu f'(\xi_c) \\
      \mu f'(\xi_c) & \mu^2 f(\xi_c) \end{array} \right)
- \left( \begin{array}{cc} d''(\xi_c) &0 \\ 0 & 0 \end{array} \right)$.
We let $\vec y_k = B \vec x_k$ and
$\bar {\vec y}_k = {B^{-1}}^\top \bar {\vec x}_k$.
We note $\bar {\vec x}_k \cdot \vec x_k =
\bar {\vec y}_k \cdot \vec y_k$.  
We let 
$C = \frac{1}{\sqrt{2}}\left( \begin{array}{cc} \bar \eta_c & -\bar \xi_c \\
      \bar \eta_c & - \bar \xi_c \end{array} \right)$.
The partition function,
thus, becomes
\begin{eqnarray}
{  Z}  &= &
e^{-S(\bar \xi_c, \xi_c,
 \bar \eta_c, \eta_c )}
\lim_{M \to \infty}
 \int
 \left[ \prod_{k=1}^M  \frac{ d \vec y_k }{2 \pi}
\right]
 \left[ \prod_{k=1}^M  \frac{ d \bar {\vec y}_k  }{2 \pi}
\right]
e^{-\frac{1}{2} \sum_{k=1}^M 
[
{ y_1}_k^2 + {y_2}_k^2 
- 2 i \bar {\vec y}_k \cdot \vec y_k
]
}
\nonumber \\ &&
\times 
e^{\frac{1}{2} \epsilon \sum_{k,l=1}^M 
[\bar {\vec y}_k ^\top B C^\top
C B^\top \bar {\vec y}_l
\frac{M'_{kl}}{\bar \eta_c^2 + \bar \xi_c^2}
- \delta_{kl} 
\bar {\vec y}_k^\top B
 B^\top \bar {\vec y}_l]
}
\nonumber \\ 
&=&
e^{-S(\bar \xi_c, \xi_c,
 \bar \eta_c, \eta_c )}
\lim_{M \to \infty}
 \int
 \left[ \prod_{k=1}^M  \frac{ d \bar {\vec y}_k  }{2 \pi}
\right]
e^{
-\frac{1}{2} \sum_{k=1}^M 
[{ \bar y}_{1k}^2 + {\bar y}_{2k}^2 ]
}
\nonumber \\ && \times
e^{
\frac{1}{2} \epsilon \sum_{k,l=1}^M 
[\bar {\vec y}_k^\top B C^\top
C B^\top \bar {\vec y}_l
\frac{M'_{kl}}{\bar \eta_c^2 + \bar \xi_c^2}
- \delta_{kl} 
 \bar {\vec y}_k^\top B
 B^\top \bar {\vec y}_l]
}
\nonumber \\ &=&
e^{-S(\bar \xi_c, \xi_c,
 \bar \eta_c, \eta_c )}
\lim_{M \to \infty}
 \int
 \left[ \prod_{k=1}^M  \frac{ d \bar {\vec y}_k  }{2 \pi}
\right]
e^{
-\frac{1}{2} \sum_{k,l=1}^M 
{\bar {\vec y}_k}^\top F'_{kl} \bar {\vec y}_l
}
\nonumber \\ &=&
e^{-S(\bar \xi_c, \xi_c,
 \bar \eta_c, \eta_c )}
(\det F'')^{-1/2}
\label{80}
\end{eqnarray}
where
\begin{eqnarray}
F''_{kl} &= &
\delta_{kl} I -
   \epsilon \frac{M'_{kl}}{\bar \eta_c^2 + \bar \xi_c^2}
  B C^\top C B^\top
+ \epsilon \delta_{kl} B B^\top
\nonumber \\ &=&
F'''_{kl} + \epsilon \delta_{kl} B B^\top
\label{81}
\end{eqnarray}
We note 
\begin{eqnarray}
\det F'' &=& (\det \left[I +  
\epsilon B B^\top I - \epsilon \frac{M'_{kl}}{\bar \eta_c^2 + \bar \xi_c^2}
  B C^\top C B^\top \right])
\nonumber \\ &=&
 (\det [I + \epsilon B B^\top ])
 \left(\det \left[ I - \epsilon \frac{M'_{kl}}{\bar \eta_c^2 + \bar \xi_c^2}
B C^\top C B^\top
(I + \epsilon B B^\top)^{-1} \right] \right)
\nonumber \\ &\sim&
 (\det F''') [\det (\delta_{kl}  + \epsilon B B^\top  \delta_{kl})]
\nonumber \\ &=&
 (\det F''') e^{ {\rm Tr} \ln 
( \delta_{kl}  + \epsilon B B^\top  \delta_{kl}) }
\nonumber \\ &\sim&
 (\det F''') e^{ {\rm Tr} 
 \epsilon B B^\top  \delta_{kl} }
\nonumber \\ &=&
 (\det F''') e^{  t {\rm Tr}  B B^\top}
\nonumber \\ &=&
 (\det F''') e^{  t {\rm Tr} B^\top B}
\nonumber \\ &=&
 (\det F''') e^{  t e^{\mu (\eta_c -1)} [f''(\xi_c) + \mu^2 f(\xi_c)] 
-t d''(\xi_c)}
\label{81a}
\end{eqnarray}
To evaluate $\det F'''$ we again use Fourier space.  We note
\begin{eqnarray}
\hat F'''(k) =
 I - B C^\top C B^\top \frac{4 \epsilon^2}{b'} 
\frac{1}{4 {b'}^2 \epsilon^2 + k^2}
\label{81b}
\end{eqnarray}
where 
\begin{eqnarray}
b' &=& [\bar \eta_c^2 + \bar \xi_c^2]^{1/2}
\nonumber \\ &=&
\frac{
\mu e^{-\mu} e^{ \mu \eta_c} f(\xi_c)
}
{\sqrt{1-\xi_c^2}}
\nonumber \\ &=&
\frac{
\mu e^{-\mu} e^{ \mu \sqrt{1-\xi_c^2}} f(\xi_c)
}
{\sqrt{1-\xi_c^2}}
\label{81bb}
\end{eqnarray}
We find
\begin{eqnarray}
\det F'''  &=& \prod_k \det \hat F'''(k)
\nonumber \\ &=&
\prod_k \det
\left(
 I - B C^\top C B^\top \frac{4 \epsilon^2}{b'} 
\frac{1}{4 {b'}^2 \epsilon^2 + k^2}
\right)
\nonumber \\
&\sim&
\prod_k 
\left(
 1 - \frac{4 \epsilon^2 a'}{b'}
\frac{1}{4 {b'}^2 \epsilon^2 + k^2}
\right)
{\rm ~as~} t \to \infty
\label{81c}
\end{eqnarray}
where 
\begin{eqnarray}
a' &=& {\rm Tr} B C^\top C B^\top 
= {\rm Tr} C B^\top B C^\top
\nonumber \\ &=&
 e^{ \mu (\eta_c-1)} 
\left[ \bar \eta_c^2 [f''(\xi_c) - d''(\xi_c)  e^{ \mu (1-\eta_c)}]
- 2 \bar \eta_c \bar \xi_c \mu f'(\xi_c) + 
\bar \xi_c^2 \mu^2 f(\xi_c) \right]
\nonumber \\ &=&
 e^{ \mu (\eta_c-1)}  {b'}^2
\left[  \eta_c^2 [f''(\xi_c) - d''(\xi_c)  e^{ \mu (1-\eta_c)}]
- 2  \eta_c  \xi_c \mu f'(\xi_c) + 
 \xi_c^2 \mu^2 f(\xi_c) \right]
\nonumber \\ &=&
 {b'}^2 c'
\nonumber \\ 
c' &=&
 e^{ -\mu } 
 e^{ \mu \sqrt{1-\xi_c^2}} 
\bigg[  (1-\xi_c^2) [f''(\xi_c) - e^{ \mu }
 e^{ -\mu \sqrt{1-\xi_c^2}} d''(\xi_c) ]
\nonumber \\ &&
- 2 \mu  \sqrt{1-\xi_c^2}  \xi_c f'(\xi_c) + 
\mu^2 \xi_c^2 f(\xi_c) \bigg]
\nonumber \\ 
\label{81cc}
\end{eqnarray}
From Eq.\ (\ref{42c}), we find
\begin{eqnarray}
\det F''' &=& 
\frac{
\left( 
\sinh t \sqrt{{b'}^2 - a'/b'}
\right)^2
}{
\left( 
\sinh b' t
\right)^2
}
\nonumber \\ &\sim&
e^{-2 t \left[
b' - [{b'}^2  - a'/b']^{1/2}
\right] 
}
\label{82}
\end{eqnarray}
Using Eqs.\ 
(\ref{55}),
(\ref{59}),
(\ref{60}),
(\ref{61}),
(\ref{61a}),
(\ref{62}),
(\ref{80}),
(\ref{81a}),
and
(\ref{82})
we find the mean excess replication rate at long times becomes
\begin{eqnarray}
\frac{\ln Z}{t N} &=&
e^{\mu (\eta_c-1)} f(\xi_c) - d(\xi_c) 
-\bar \xi_c \xi_c - \bar \eta_c \eta_c
+ \sqrt{\bar \eta_c^2 + \bar \xi_c^2}
\nonumber \\ && 
+ \frac{1}{N} 
b' \left [
1 - \left[ 1 - a'/{b'}^3    \right]^{1/2}
\right]
\nonumber \\
\label{83}
\end{eqnarray}

\section*{Appendix E}
In this Appendix, we evaluate the functional integral expression
of the full probability distribution function for the
parallel model.
Here ${ Z}$ has the form
\begin{eqnarray}
{ Z} &=& \lim_{M \to \infty}
 \int \left[ \prod_{k=1}^M  \frac{i \epsilon N d \bar \xi_k d \xi_k }{2 \pi}
\right]
e^{\epsilon \sum_{k=1}^M 
[ N f(\xi_k) - N \bar \xi_k \xi_k - N \mu + \Delta f] }
\nonumber \\ &&
\times
 \int \left[ \prod_{k=0}^M \prod_{j=1}^N
\frac{d \vec z_k^*(j) d \vec z_k(j)}{\pi^2} \right]
e^{- \sum_{k,l=0}^M \vec z_k^*(j) S_{kl}(j) \vec z_l(j) }
\nonumber \\ && \times e^{\sum_{j=1}^N [\vec J^*(j) \cdot \vec z_M(j) +
              \vec J_0(j) \cdot \vec z_0^*(j)] }
\nonumber \\ &=&
 \int \left[ \prod_{k=1}^M  \frac{i \epsilon N d \bar \xi_k d \xi_k }{2 \pi}
\right]
e^{\epsilon \sum_{k=1}^M 
[ N f(\xi_k) - N \bar \xi_k \xi_k - N \mu + \Delta f] }
\nonumber \\ &&
\times e^{\sum_{j=1}^N J_0(j) S_{0M}^{-1}(j) J^*(j)}
\prod_{j=1}^N [\det S(j)]^{-1} 
\label{65}
\end{eqnarray}
where the matrix $S(j)$ has the form
\begin{eqnarray}
S(j) = 
\left( \begin{array}{ccccc} 
I & 0 & 0 &\ldots & 0\\
- A_1(j) & I & 0 & \ldots &0 \\
0 & - A_2(j) & I & \ldots &0\\
&&&\ddots\\
0&& \ldots & -A_M(j) &  I
\end{array} \right) 
\label{66}
\end{eqnarray}
Equation (\ref{64}) evaluates as
\begin{eqnarray}
P(\{ \vec n \}, t) 
&=& \lim_{M \to \infty}
 \int \left[ \prod_{k=1}^M  \frac{i \epsilon N d \bar \xi_k d \xi_k }{2 \pi}
\right]
e^{\epsilon \sum_{k=1}^M 
[ N f(\xi_k) - N \bar \xi_k \xi_k - N \mu + \Delta f] }
\nonumber \\ &&
\times
\sum_{\{ \vec n_0 \} }
P(\{ \vec n_0 \})
\prod_{j=1}^N \{ [Q(j)]_{\alpha_0(j) \alpha(j)} \}
\label{67}
\end{eqnarray}
where
\begin{eqnarray}
Q(j) &=& 
\prod_{k=1}^M [I + \epsilon \mu \sigma_1 + \epsilon \bar \xi_k \sigma_3 ]
\nonumber \\ &\sim&
\hat T e^{\epsilon \sum_{k=1}^M
[ \mu \sigma_1 + \bar \xi_k \sigma_3 ] }
\nonumber \\ &=&
 \hat T e^{ \int_0^t  dt'
[ \mu \sigma_1 + \bar \xi(t') \sigma_3 ] }
\label{68}
\end{eqnarray}

\section*{Appendix F}

In this Appendix, we evaluate the expressions necessary
to determine the parallel model distribution function in the
Gaussian central region.
We define $u_* = \lim_{N \to \infty} \langle u \rangle$ and
$ u = u_* + \delta u$.  We note that
$\xi_k = \xi(t') $ satisfies
\begin{eqnarray}
\xi(t') &=& u_*, t'=0
 \nonumber \\
\xi(t') &=& \xi_c,   1 \ll t'  \ll t
 \nonumber \\
\xi(t') &=& u, t' = t
\label{211a}
\end{eqnarray}
We note that to $O(\delta u^2)$ 
\begin{eqnarray}
\xi(u_* + \delta u, t') = \xi(u_*, t' + \delta t)
\label{213}
\end{eqnarray}
in the range $1 \ll t'$, where
\begin{eqnarray}
\delta t =  - \frac{ \delta u}{2 \mu u_*} + 
\frac{\mu - f'(u_*)/u_*}{(2 \mu u_*)^2} \delta u^2
\label{214}
\end{eqnarray}
because  the differential equation
\begin{eqnarray}
\frac{d \xi(t')}{d t'} =
\mu (1+u) \langle \sigma_2(k) \rangle_+
+ \mu (1-u) \langle \sigma_2(k) \rangle_-
\label{212}
\end{eqnarray}
is invariant under this shift, and the
boundary condition $\xi(t) = u_* + \delta u$
is satisfied to $O(\delta u^2)$
with the chosen value of $\delta t$, Eq.\ (\ref{214}).
We now evaluate Eq.\ (\ref{210}) to $O(\delta u^2)$.
Using the shift property (\ref{213}) and 
conditions (\ref{211a}), 
we find to $O(\delta u^2)$
\begin{eqnarray}
\int_0^t [  f(t') &-&  \bar \xi(t') \xi(t') -  \mu ] dt'
= ({\rm const}) 
-\delta t f(\xi_c) + \delta t \frac{
 f(u_*)
+
f(u_* + \delta u) 
}{2}
\nonumber \\ &&
+\delta t [\xi_c f'(\xi_c)] - \delta t \frac{u_* f'(u_*) + 
(u_* + \delta u)f'(u_* + \delta u_*)}{2}
\label{215}
\end{eqnarray}
and
\begin{eqnarray}
- \frac{1+u}{2} \ln \frac{1+u}{2} 
- \frac{1-u}{2} \ln \frac{1-u}{2} 
&= &
- \frac{1+u_*}{2} \ln \frac{1+u_*}{2} 
- \frac{1-u_*}{2} \ln \frac{1-u_*}{2} 
\nonumber \\ &&
- \delta u \ln \frac{1+u_*}{1-u_*} -
\frac{\delta u^2}{2} \frac{1}{1-u_*^2}
\label{215a}
\end{eqnarray}
Using the shift property (\ref{213}) and boundary
conditions (\ref{211a}), we find to $O(\delta u^2)$
\begin{eqnarray}
\frac{[Q(u_* + \delta u)]_{+}}{Q_*} &=&
e^{-\lambda^* \delta t}
\bigg[
\frac{1+u_*}{2}  +
\delta t
\left(
\frac{1-u_*}{2}\mu  +
\frac{1+u_*}{2} 
f'(u_*)
\right)
\nonumber \\ && + \frac{\delta t^2}{2}
\frac{1+u_*}{2} 
\left(
- 2 \mu u_* f''(u_*) + \mu^2 + f'(u_*)^2 
\right)
\bigg]
\label{216}
\end{eqnarray}
and
\begin{eqnarray}
\frac{[Q(u_* + \delta u)]_{-}}{Q_*} &=&
e^{-\lambda^* \delta t}
\bigg[
\frac{1-u_*}{2}  +
\delta t
\left(
\frac{1+u_*}{2}\mu  -
\frac{1-u_*}{2} 
f'(u_*)
\right)
\nonumber \\ &&
+ \frac{\delta t^2}{2}
\frac{1-u_*}{2} 
\left(
2 \mu u_* f''(u_*)  + \mu^2 + f'(u_*)^2 
\right)
\bigg]
\label{217}
\end{eqnarray}
where $\lambda_* = \sqrt{\mu^2 + \bar \xi_c^2}$.

\section*{Appendix G}
In this Appendix, we evaluate the expressions necessary
to determine the Eigen model distribution function in the
Gaussian central region.
 We note that to $O( \delta t^2)$
\begin{eqnarray}
\xi(u_* + \delta u, t') &=& \xi(u_*, t' + \delta t)
\nonumber \\ 
\eta(u_* + \delta u, t') &=& \eta(u_*, t' + \delta t)
\label{245}
\end{eqnarray}
in the range $1 \ll t'$, where
\begin{eqnarray}
\delta t =  - \frac{ \delta u}{2 \mu u_* f(u_*)} + 
\frac{d'(u_*)  - f'(u_*) + \mu u_* f(u_*) + \mu u_*^2 d'(u_*) }
{u_* [2 \mu u_* f(u_*) ]^2} \delta u^2
\label{246}
\end{eqnarray}
because  the differential equation
\begin{eqnarray}
\frac{d \xi(t')}{d t'} =
\bar \eta(t') (1+u) \langle \sigma_2(k) \rangle_+
+ \bar \eta(t') (1-u) \langle \sigma_2(k) \rangle_-
\label{247}
\end{eqnarray}
is invariant under this shift, and the
boundary condition $\xi(t) = u_* + \delta u$
is satisfied to $O(\delta u^2)$
with the chosen value of $\delta t$, Eq.\ (\ref{246}).

Using the shift property  (\ref{245}) 
we find to $O(\delta u^2)$
\begin{eqnarray}
\int_0^t  &&
[ e^{-\mu + \mu \eta(t')} f(\xi(t')) - d(\xi(t')) ]
 dt'
\nonumber \\ &=&
 ({\rm const}) 
-\delta t [e^{-\mu + \mu \eta_c}  f(\xi_c) -d(\xi_c)]
 + \delta t [ f(u_*) - d(u_*)] 
\nonumber \\ &&
+ \frac{ \delta t \delta u}{2} 
\left[ f'(u_*) - d'(u_*) + 
\mu f(u_*) \left. \frac{d \eta}{d u}\right\vert_{*,t} \right]
\nonumber \\ &=& ({\rm const}) + O(\delta u^3)
\label{248}
\end{eqnarray}
where
\begin{eqnarray}
\left. \frac{d \eta}{d u}\right\vert_{*,t} 
&=&
\left. \frac{d }{d u}\right\vert_{*,t} 
\left( 
\frac{1+u}{2} \frac{Q_-}{Q_+}
+
\frac{1-u}{2} \frac{Q_+}{Q_-}
\right)
\nonumber \\ &=&
\frac{d'(u_*) - f'(u_*)}{\mu f(u_*) }
\label{249}
\end{eqnarray}
We also find
\begin{eqnarray}
\int_0^t && [
- \bar \xi(t') \xi(t') - \bar \eta(t') \eta(t') ]
\nonumber \\ &=&
\delta t [\bar \xi_c \xi_c + \bar \eta_c \eta_c]
- \delta t [\mu f(u_*) + u_* f'(u_*)- u_* d'(u_*)  ]
\nonumber \\  &&
- \frac{ \delta t \delta u}{2} 
\left. \frac{d \eta}{d u}\right\vert_{*,t} 
\left[ 
\mu u_* f'(u_*) + \mu^2 f(u_*) + \mu f(u_*)
\right]
\nonumber \\  &&
- \frac{ \delta t \delta u}{2} 
\left[ 
f'(u_*) - d'(u_*) + u_* f''(u_*)  - u_* d''(u_*)
+ \mu f'(u_*)
\right]
\label{250}
\end{eqnarray}
We also find
\begin{eqnarray}
\frac{[Q(u_* + \delta u)]_{\alpha}}{Q_*} &=&
e^{-\lambda^* \delta t}
\bigg[
\frac{1+\alpha u_*}{2}  +
\delta t
\bigg(
\frac{1-\alpha u_*}{2}\mu f(u_*) 
\nonumber \\ && +
\alpha \frac{1+\alpha u_*}{2} 
[f'(u_*) - d'(u_*)]
\bigg)
\nonumber \\ && + \frac{\delta t^2}{2}
\bigg(
\frac{1-\alpha u_*}{2} 
\left. \frac{d \bar \eta}{d t'}\right\vert_{*,t} 
+ \alpha \frac{1+\alpha u_*}{2}
\left. \frac{d \bar \xi}{d t'}\right\vert_{*,t} 
\nonumber \\ &&+
\frac{1+\alpha u_*}{2} 
\left[
\mu^2 f(u_*)^2  + (f'(u_*)   - d'(u_*))^2 
\right]
\bigg)
\bigg]
\label{251}
\end{eqnarray}
where
\begin{eqnarray}
\lambda^* &=& \bar \xi_c \xi_c + \bar \eta_c \eta_c
\nonumber \\
\left. \frac{d \bar \eta}{d t'}\right\vert_{*,t} 
&=&
- 2 \mu u_* f(u_*) \left[
\mu f'(u_*) 
- \mu f'(u_*) + \mu d'(u_*)
\right]
\nonumber \\ 
\left. \frac{d \bar \xi}{d t'}\right\vert_{*,t} 
&=&
- 2 \mu u_* f(u_*) \left[ f''(u_*) - d''(u_*)
+ \mu f'(u_*) \left. \frac{d \eta}{d u}\right\vert_{*,t} 
\right]
\label{252}
\end{eqnarray}
We also have Eq.\ (\ref{215a}).

\end{document}